\documentclass[letterpaper]{JHEP3}
\usepackage{axodraw}
\usepackage{amsmath,amssymb}
\usepackage{amscd}

\newcommand{\la}{\langle}
\newcommand{\ra}{\rangle}

\newcommand{\beqn}{\begin{eqnarray}}
\newcommand{\eeqn}{\end{eqnarray}}
\newcommand{\nn}{\nonumber}
\newcommand{\gs}{{\mathrm{g}}_s}
\newcommand{\as}{\alpha_s}

\newcommand{\mi}{\mathrm{i}}
\newcommand{\dd}{\displaystyle}
\def\MS{$\overline{\mathrm{MS}}$\ }

\def\s#1#2{s_{#1#2}}
\def\bom#1{{\mathbf{#1}}}
\def\bmi#1{{\mbox{\boldmath ${#1}$}}}

\def\cm{M}
\def\ket#1{\big|{#1}\big\ra}
\def\bra#1{\big\la{#1}\big|}

\newbox\charbox
\newbox\slabox
\def\s#1{{      
        \setbox\charbox=\hbox{$#1$}
        \setbox\slabox=\hbox{$/$}
        \dimen\charbox=\ht\slabox
        \advance\dimen\charbox by -\dp\slabox
        \advance\dimen\charbox by -\ht\charbox
        \advance\dimen\charbox by \dp\charbox
        \divide\dimen\charbox by 2
        \raise-\dimen\charbox\hbox to \wd\charbox{\hss/\hss}
        \llap{$#1$}
}}
\def\ArrowGluon(#1,#2)(#3,#4)#5#6{
\put(\axoxoff,\axoyoff){
}
}

\def\ArrowGluonn(#1,#2)(#3,#4)#5#6{
\put(\axoxoff,\axoyoff){
}
}
\title{General subtraction method for numerical calculation of
  one-loop QCD matrix elements}

\author{Zolt\'an Nagy and Davison E. Soper\\
Institute of Theoretical Science\\
University of Oregon\\
Eugene, OR  97403-5203, USA\\
E-mail: \email{nagyz@physics.uoregon.edu},
\email{soper@physics.uoregon.edu}
}

\abstract{
We present a subtraction scheme for eliminating the ultraviolet, soft, and
collinear divergences in the numerical calculation of an arbitrary
one-loop QCD amplitude with an arbitrary number of external legs.
The subtractions consist of local counter terms in the space of the
four-dimensional loop momentum. The ultraviolet subtraction terms
reproduce ${\overline {\rm MS}}$ renormalization. The key point in the
method for the soft and collinear subtractions is that, although the
subtraction terms are defined graph-by-graph and the matrix element is
also calculated graph-by-graph, the sum over graphs of the integral of
each the subtraction term can be evaluated analytically and provides
the well known simple pole structure that arises from subtractions from
real emission graphs, but with the opposite sign.
}

\preprint{hep-ph/0308127}

\begin{document}

\section{Introduction}
\label{sec:intro}

Tests of the Standard Model or one of its extensions frequently involve
high momentum transfer processes in which one is looking for effects from
new interactions or particles. In most cases, at least some of the
particles interact via the strong interactions. Then, in order to obtain
reasonably accurate predictions for the expected cross sections, it is
necessary to calculate the cross section at next-to-leading order (NLO)
in quantum chromodynamics (QCD). Sometimes this is possible using the
currently available theoretical tools, sometimes not. This paper concerns
a method for potentially extending the range of problems for which a
next-to-leading order calculation is possible.

An NLO calculation necessarily involves a virtual loop integration. That
is, one has an integral
\beqn
\int\! \frac{d^{d} l}{(2\pi)^{d}}\,
\widetilde\Gamma(k_1,\dots,k_n;l)
\label{loop}
\eeqn
over a loop momentum $l$ in $d = 4 - 2\epsilon$ dimensions, with momenta
$k_1,\cdots,k_n$ leaving the graph at vertices around the loop. Of
course, one wants $d = 4$ in the end, but there can be infrared
divergences, which are temporarily regulated by working with $d \ne 4$.
(There can also be ultraviolet divergences, but we ignore these for this
initial discussion.) The infrared divergences arise because we work with
a gauge theory containing massless particles. When some or all of the
$k_i$ are lightlike, the integral for $d = 4$ can have divergences
arising from regions of $l$-space in which the momentum of two of the
propagators in the loop approaches a line parallel to one of the external
momenta or in which the momentum of one of the propagators in the loop
approaches zero. This leads to poles of the form $1/\epsilon^2$ and
$1/\epsilon$ in the integral.

The traditional method of dealing with integrals like (\ref{loop}),
initiated in the context of collider physics cross sections by Ellis,
Ross, and Terrano \cite{Ellis:1981wv}, is to calculate the integrals
analytically. The result is expressed in the form of the residues of the
$1/\epsilon^2$ and $1/\epsilon$ poles and a remaining finite piece, which
contains the most important physical information.  It would seem not to
be helpful to have a result with $1/\epsilon^2$ and $1/\epsilon$ terms,
but it is helpful. There are other integrations involved in the
calculation of a physical observable. These are integrations over the
momenta of final state quarks and gluons in Feynman graphs without
virtual loops. As long as the observable has the property known as
infrared safety, the integration over final state momenta will produce 
$1/\epsilon^2$ and $1/\epsilon$ poles that precisely cancel the poles
from the virtual loop integrals. The integrations over final state
momenta are generally much too complicated to perform analytically. One
can subtract from the integrand certain simple integrands that match the
complete integrand in one of the soft or collinear limits that gives a
divergence. Then the difference is integrable in $d = 4$ dimensions and
this integral is performed numerically. The integral of each subtraction
term is added back, but this time with the integration performed
analytically. This gives $1/\epsilon^2$ and $1/\epsilon$ poles that cancel
against the poles from the virtual loop integrals.

In order to make this method work, one needs analytical calculations of
the virtual loop integrals that appear. Ellis, Ross, and Terrano
\cite{Ellis:1981wv} supplied the one-loop graphs with $n=2$, $n=3$, and
$n=4$ external legs that occur in a calculation for $e^+ + e^- \to 3\
{\it jets}$. Ellis and Sexton \cite{Ellis:1986er} supplied the additional
$n=2$, $n=3$, and $n=4$ graphs that are needed for $p + \bar p \to 2\ {\it
jets}$. More  recently, there have been heroic calculations by Bern,
Dixon, and Kosower \cite{Bern:1993mq,Bern:1995fz,Bern:1997ka,Bern:1998sc}
and by Kunszt, Signer, and Tr\'ocsayni \cite{Kunszt:1994tq} for $n=5$.
With these results, one can calculate, for instance, $e^+ + e^- \to 4\
{\it jets}$ and $p + \bar p \to 3\ {\it jets}$. There are also
calculations available with massive particles, such as massive quarks.

One wonders if we could not get beyond $n=5$ by simply performing the
virtual loop integrals numerically. Not only might one be able to get to
diagrams with more external legs, but also one could have the flexibility
to change the lagrangian, add masses, make approximations, and so forth
by simply altering the integrand. Of course, one is not going to perform
integrals in $d = 4 - 2\epsilon$ dimensions numerically. Instead, one
would have to do something about the infrared divergences first, then
take $d=4$.

Something like this was tried in 
Ref.~\cite{Soper:1998ye,Soper:1999xk,Soper:2001hu,Kramer:2002cd} in the
case of $e^+ + e^- \to 3\ {\it jets}$. There, the approach was to
perform the integral over the energy $l^0$ in Eq.~(\ref{loop})
analytically and then to put all of the integrals over loop momenta $\vec
l$ and final state particle momenta $\vec p$ together, summing over
different contributions to the observable inside the integration. Then
the real-virtual cancellations alluded to above happen inside the
integrals, so that the integrals are convergent in three space
dimensions. All of the integrals can be performed at once by numerical
integration.

In this paper, we present a different proposal, one that is very close to
the traditional method. One would subtract from the integrand (\ref{loop})
certain simple integrands that match the complete integrand in one of the
soft or collinear limits that gives a divergence. Then the difference
would be integrable in $d = 4$ dimensions and this integral would be
performed numerically. The integral of each subtraction term would be
added back, but this time with the integration performed analytically.
This gives the same $1/\epsilon^2$ and $1/\epsilon$ poles that one has in
an analytic calculation, ready for cancellation as usual.

We should mention that performing the 4-dimensional loop integral
numerically is not completely without difficulties. In the massless
theory, the integrand contains singularities $1/(p^2 + \mi 0)$ that lie on
cones in the loop momentum space. One must deform the integration contour
to avoid these singularities, as indicated by the ``$+\mi 0$''
prescription. However, this same problem occurs in 
Refs.~\cite{Soper:1998ye,Soper:1999xk,Soper:2001hu,Kramer:2002cd}, where
it is handled in a straightforward manner.

The issue addressed in this paper is to find what one should
subtract from the integrand. We treat this problem in QCD with massless
quarks, leaving the problem of other lagrangians and non-zero masses for
future work. We find that there is a set of subtractions that is quite
simple in the sense that the subtraction terms can be easily generated by
the same sort of computer algebra that generates the original integrand
in (\ref{loop}). Furthermore, adding back the subtraction terms is
straightforward: the residues of the $1/\epsilon^2$ and $1/\epsilon$ poles
are the familiar ones, while the finite term that comes along with the
poles is simple.

There has been other recent work aimed at automating the calculation of
loop integrals in such a way that those integrals that are not amenable
to analytic evaluation could be evaluated numerically
\cite{Binoth:2000ps,Binoth:2002xh,Binoth:2003ak}. This approach
makes use of the Feynman parameter representation of the diagrams and is
quite ambitious in that it applies to multiloop diagrams. In the
present paper we deal only with one-loop diagrams, although with an
arbitrary number of external legs. Our treatment is then to subtract from
the integrands certain terms that remove divergences, but to otherwise
leave the integrand entirely as given by the momentum space Feynman rules.
Thus we aim at maximal simplicity of the method.

\section{\label{sec:strategy}Strategy}

We seek a method for performing subtractions on one-loop graphs in
such a way that we are left with an integral that can be performed
numerically in four dimensions. That means that we must take care of
ultraviolet, collinear, and soft divergences.

We note first of all that in the case of tree graphs it is common to sum
all of the graphs that contribute to a given set of external particles.
Many graphs may contribute, but one can sum these graphs algebraically
before entering into a numerical computation. With loop graphs, the
situation is different. If we are going to integrate over the loop
momentum numerically, we (or our computers) have to know the structure of
the integrand, which is very different for different graphs. Thus we will
work graph by graph.

Working graph by graph, the choice of gauge matters. We take the simplest
choice, the Feynman gauge. The external lines in each graph represent
physical, on-shell particles. Thus our algorithms are based on having
spinors $u(p,s)$, $\bar u(p,s)$, $v(p,s)$ and $\bar v(p,s)$ for
external quarks and antiquarks and polarization vectors
$\varepsilon^\mu(p,s)$ for external gluons. The polarization vectors can be
defined using any physical gauge one cares to use. Our analysis makes use
of the relation $p\cdot\varepsilon(p,s) = 0$. If it were desired, one
could use unphysical gluon polarizations, with $\sum_s \varepsilon^\mu(p,s)
\varepsilon^\nu(p,s)^* \longrightarrow - g^{\mu\nu}$. Then the relation 
$p\cdot\varepsilon(p,s) = 0$ fails, and the reader is asked to
substitute the following argument. We calculate a one-loop amplitude,
which must be multiplied by a tree level amplitude to calculate a term in
a cross section. Where $p_\mu$ appeared contracted with a polarization
vector, it now appears contracted with the tree level amplitude. But 
the contraction of $p_\mu$ with the tree level amplitude vanishes.

Thus our aim is to examine one-loop graphs with on-shell massless
external particles. For each such graph, there may be ultraviolet,
soft, and collinear divergences. In Secs.~\ref{sec:UV},
\ref{sec:soft}, and \ref{sec:collinear}, respectively, we will define
subtractions for the integrand that eliminate each divergence. The
ultraviolet subtractions are required for modified minimal subtraction
($\overline{\mathrm{MS}}$) renormalization. All that we have to do is
to express the \MS counterterm as a subtraction on the integrand. In
the case of the soft and collinear subtractions, we do not want to
change the value of the graph, so we add back what we subtracted.
However, we subtract something from the integrand and add back its
integral summed over graphs. What we need to add back is {\it not}
simple graph by graph. However, when we sum over graphs we get a
simple prescription for what to add back. We begin our investigations
in the next section by setting up some useful notation.

\section{\label{sec:graphs}Matrix elements at tree and one-loop level}

In this section, we set up our notation. First, we define a notation for
the color and spin structure of matrix elements, following in general the
notation of Catani and Seymour \cite{Catani:1997vz}. Then we add some
notation specific to one-loop graphs. Finally, we define the wave
function renormalization factors needed to relate amputated Green
functions to the $S$-matrix.

The matrix element with $m$ external QCD partons has the following
general structure:
\beqn
\cm^{c_1,\dots,c_m; s_1,\dots,s_m}(p_1,\dots,p_m),
\eeqn
where $\{c_1,\dots,c_m\}$, $\{s_1,\dots,s_m\}$ and $\{p_1,\dots,p_m\}$ are
respectively color indices (which take $N_c^2 - 1$ values for each gluon
and $N_c$ values for each quark or anti-quark), spin indices (which take
$d - 2$ values for gluons in $d$-dimensional spacetime, and
$2$ values for quarks and antiquarks) and momenta. We take the momenta
$p_i$ to be coming out of the graph, so that $p_i^2 = 0$ with $p_i^0 > 0$
for outgoing partons and $p_i^0 < 0$ for incoming partons. The matrix
element is computed by contracting amputated Green functions with $\bar
u(p,s)$ for an outgoing quark, $v(p,s)$ for an outgoing antiquark, and
$\varepsilon(p,s)^*$ for an outgoing gluon. For incoming partons, the
construction is analogous, using $u(-p,s), \bar v(-p,s)$ and
$\varepsilon(-p,s)$. (There are identities available
\cite{Gunion:1985vc,Mangano:1991by} to simplify the spin wave functions,
but these do not concern us in this paper.)

It is useful to represent the matrix element as a vector $\big|{\cal
M}\big\ra$ in color $\otimes$ spin space. One introduces a basis
$\{\ket{c_1,\dots,c_m} \otimes \ket{s_1,\dots,s_m} \}$ in color $\otimes$
spin space in such a way that
\beqn\label{cmmdef}
\cm^{c_1,\dots,c_m; s_1,\dots,s_m}(p_1,\dots,p_m) \equiv
\Bigl( \bra{c_1,\dots,c_m} \otimes \bra{s_1,\dots,s_m} \Bigr)
\big|{\cal M}(p_1,\dots,p_m)\big\ra \;.
\eeqn
With this notation, the matrix element squared (summed over final-state
colors and spins) $|M|^2$ can be written as
\beqn
|M|^2 = \left\la {\cal M}\left(p_1,\dots, p_m\right)\right|
\left.{\cal M}\left(p_1,\dots, p_m\right)\right\ra .
\eeqn

\subsection{\label{subsec:color}Color space}

Following the notation of \cite{Catani:1997vz}, we associate a color
charge operator ${\bom T}_i$ with the emission of a gluon from each
parton $i$. Its action onto the color space is defined by
\beqn
{\bom T}_i = 
T_{c b}^a \ket{a} \ket{c}_{\!i}\, {\vphantom{\big\rangle}}_i\! \bra{b}
\;\;,
\eeqn
where $T_{c b}^a$ is the appropriate color matrix ($i f_{cab}$ if the
emitting particle $i$ is a gluon, $t^a_{\alpha \beta}$ if the emitting
particle $i$ is a quark and ${\bar t}^a_{\alpha \beta} = - t^a_{\beta
\alpha }$ if the emitting particle $i$ is an antiquark). That is,
\beqn
\bra{c_1,\dots, c_i,\dots,c_m, c} {\bom T}_i
\ket{b_1,\dots, b_i,\dots,b_m} = \delta_{c_1 b_1} \cdots
T_{c_i b_i}^c \cdots\delta_{c_m b_m} \;\;.
\eeqn
One also defines the corresponding adjoint operators ${\bom T}_i^\dagger$
for absorbing a gluon onto parton $i$. Then one defines a dot product of
operators, ${\bom T}_i \cdot {\bom T}_j \equiv  {\bom T}_i^\dagger {\bom
T}_j$ for emitting a gluon from line $i$ and absorbing it on line $j$.
For $i \ne j$ this is
\beqn
{\bom T}_i \cdot {\bom T}_j = {\bom T}_j \cdot {\bom T}_i = 
T^a_{c_i b_i}T^a_{c_j b_j} 
\ket{c_i}_{\!i}\,\ket{c_j}_{\!j}\ 
{\vphantom{\big\rangle}}_i\!\bra{b_i}
{\vphantom{\big\rangle}}_j\!\bra{b_j}\ \;\;.
\eeqn
The case $i=j$ is special:
\beqn
{\bom T}_i \cdot {\bom T}_i = C_i\;\;,
\eeqn
where $C_i$ is the Casimir operator, that is, $C_i=C_A=N_c$ if $i$ is a
gluon and $C_i=C_F=(N_c^2-1)/2N_c$ if $i$ is a quark or anti-quark.

Note that, by definition, each vector $\big|{\cal M}\big\ra$ is
a color singlet. Therefore color conservation is simply
\beqn \label{cocon}
\sum_{i=1}^m {\bom T}_i \;\big|{\cal M}(p_1,\dots,p_m)\big\ra = 0 \;.
\eeqn

Using this notation, we also define the square of color correlated
amplitudes, $|{\cal{M}}^{i,k}|^2$, as follows
\beqn
\label{colam}
|{\cal{M}}^{i,k}|^2 &\equiv&
\big\la{\cal M}(p_1,\dots,p_m)\big| \,{\bom T}_i 
\cdot {\bom T}_k \,\big|{\cal M}(p_1,\dots,p_m)\big\ra
\nonumber \\
&=&
\left[ {\cal M}^{\,a_1\dots b_i\dots b_k\dots a_m}(p_1,\dots,p_m)
\right]^*
\; T_{b_ia_i}^c \, T_{b_ka_k}^c
\; {\cal M}^{a_1\dots a_i\dots a_k\dots a_m}(p_1,\dots,p_m) \;.
\eeqn

\subsection{\label{subsec:oneloop}One-loop graphs}

As discussed in the introduction of this section, we wish to calculate
the one-loop, $m$-parton matrix element graph by graph using numerical
integration. In this subsection, we define some tools for this
calculation. We use the vector notation in color and spin space of the
previous subsection to write the one-loop matrix element as a sum of the
one-loop Feynman graphs,
\beqn
\big|{\cal M}(p_1,\dots,p_m)\big\ra = 
\sum_{\mathrm{graphs}\,G} 
\big|{\cal G}(G;p_1,\dots,p_m)\big\ra\;\;.
\eeqn
The $|{\cal G}\ra$ vector is related to the one-loop, amputated Green
function contracted with the appropriate spinors and gluon
polarization vectors by
\beqn
{\cal G}^{c_1,\dots,c_m; s_1,\dots,s_m}(G;p_1,\dots,p_m) \equiv
\Bigl( \bra{c_1,\dots,c_m} \otimes \bra{s_1,\dots,s_m} \Bigr) 
\big|{\cal G}(G;p_1,\dots,p_m)\big\ra\;\;.
\eeqn

The graph $G$ contains a one-loop, one-particle-irreducible subgraph
$\Gamma$ with $n = n(G)$ external lines and $n$ internal lines around the
loop. Let us label momenta of the internal lines $l_1,l_2,\dots l_n$. We
integrate over the loop momentum. One can consider $l_k$ for some
specific index $k$ to be the independent loop momentum, which we call
$l$.\footnote{Alternatively, $l$ could be the average of the $l_k$. This
choice is convenient for constructing the ultraviolet counterterm for a
loop that needs ultraviolet renormalization. See Sec.~\ref{sec:UV}.}
Once we make a choice of $l_k$ as the independent variable, any of the
propagator momenta around the loop is related to $l$ by an equation of
the form
\beqn
l_i = l_k + \sum_{j=1}^m A(G,k)_{ij}\,p_j,
\label{loopspace}
\eeqn
where the matrix $A$ is determined by the structure of the graph and by
our choice $l = l_k$. We write the integration over loop momentum as
\beqn
\int\! \frac{ d^d l}{(2\pi)^d}
\;\;.
\eeqn
Since the integration measure in the space of loop momenta does not
depend on the choice of origin ({\it e.g.} $dl_{k'} = dl_k$) we omit a
label on $dl$ indicating the definition of $l$. In the actual numerical
integration, one chooses a definition for $l$, then
chooses a random value for $l$ thus defined, and then computes the $l_i$
from Eq.~(\ref{loopspace}). Thus our notation is
\beqn
\big|{\cal G}(G;p_1,\dots,p_m)\big\ra =
\int\! \frac{ d^d l}{(2\pi)^d}\
\big| \widetilde{\cal G}(G;\{l\};p_1,\dots,p_m)\big\ra
\;\;,
\eeqn
where $\widetilde {\cal G}$ denotes the integrand for graph $G$ and $\{l\}$
is a shorthand for $l_1,\dots,l_{n}$.
   
The loop integral could be ultraviolet divergent. This happens for $n =
2,3$ and for $n=4$ in the case that all of the lines entering the loop are
gluons. In this case we define a suitable UV counterterm that reproduces
the same result as $\overline {\rm  MS}$ renormalization. We define the
counter terms in Sec.~\ref{sec:UV}. 

In certain graphs, the loop integration is infrared divergent. For
instance  when an internal gluon line connects two external lines the
integral has soft, collinear, and soft-collinear poles. To handle the
infrared divergences, we introduce simple counter terms that we subtract
from the integrand for the numerical integration and then add back
analytically. We discuss this in Sec.~\ref{sec:soft} for the soft
divergences and in Sec.~\ref{sec:collinear} for the collinear
divergences.

\subsection{\label{subsec:quarkexternalleg}S-matrix}

The $S$-matrix can be constructed from the amputated Green functions 
by means of the LSZ reduction formula
\beqn 
\big|{\cal M}_{\rm full}(p_1,\dots,p_m)\big\ra =
\left(\prod_i \sqrt r_i\right)
\big| {\cal M}_{\rm full}^{(T)}(p_1,\dots,p_m)\big\ra
\;\;.
\eeqn
Here we consider both sides of the equation to be vectors in color and
spin space. The subscripts ``full'' indicate that this formula applies to
the matrix element summed over orders of perturbation theory. The $T$ (for
``truncated'') superscript on ${\cal M}$ on the right indicates that the
matrix element is calculated by multiplying amputated Green functions by
the appropriate Dirac spinors and polarization vectors. The factors
$\sqrt r_i$, one for each external particle, are of two types, one for
gluons and one for quarks and antiquarks. They are defined from the
residues of the poles of the corresponding propagators at $p^2 = 0$.
Specifically, when the flavor of particle $i$ is $q$ or $\bar q$ then
$r_i = R_{\psi}$, where the renormalized quark propagator near $p^2= 0$
has the behavior
\beqn 
iS(p) \sim R_{\psi}\frac{i\s{p}}{p^2 + \mi 0}
\hskip 1 cm p^2 \to 0\;\;.
\label{Rpsidef}
\eeqn
When the flavor of particle $i$ is $g$ then
$r_i = R_{A}$ where the renormalized gluon propagator near $p^2= 0$
has the behavior
\beqn 
iD^{\mu\nu}(p) \sim R_{A}\frac{-ig^{\mu\nu}}{p^2 + \mi 0} +\cdots
\hskip 1 cm p^2 \to 0\;\;,
\label{RAdef}
\eeqn
where the omitted terms are gauge terms proportional to $p^\mu$ or
$p^\nu$.

The constants $r_i$ have perturbative expansions
\beqn 
r_i = 1 + r^{(1)}_i + \cdots,
\eeqn
where $r^{(1)}_i \propto \alpha_s^1$.
Thus, including terms up to one-loop order,
\beqn 
\big|{\cal M}_{\rm full}(p_1,\dots,p_m)\big\ra &=&
\big| {\cal M}_{\rm tree}(p_1,\dots,p_m)\big\ra
+
\big| {\cal M}^{(T,1)}(p_1,\dots,p_m)\big\ra
\nonumber\\
&&
+ \sum_i \frac{1}{2} \,
r^{(1)}_i \,\big| {\cal M}_{\rm tree}(p_1,\dots,p_m)\big\ra
\;\;.
\label{externallegs}
\eeqn
Here $\big| {\cal M}^{(T,1)}\big\ra$ is the one-loop contribution to
$\big| {\cal M}^{(T)}\big\ra$. Elsewhere in this paper we denote $\big|
{\cal M}^{(T,1)}\big\ra$ simply as $\big| {\cal M}\big\ra$.

Now we need to evaluate the residue constants $r_i^{(1)}$. Consider
first the quark propagator. Expanding to order $\alpha_s$ we have
\beqn 
i S(p) =  \frac{\mi\s{p}}{p^2 + \mi 0}
+\frac{\mi\s{p}}{p^2 + \mi 0}\
(-\mi\Sigma_R(p))
\frac{\mi\s{p}}{p^2 + \mi 0}
\;\;,
\label{quarkprop1}
\eeqn
where $\Sigma_R(p)$ is the one-loop renormalized quark self-energy graph.
It has the form
\beqn 
\Sigma_R(p) = A(p^2)\, \s{p}
- Z^{(1)}_\psi\, \s{p}
\;\;,
\eeqn
where evaluation of the graph in $d - 4 - 2\epsilon$ dimensions gives
\beqn 
A(p^2) = - C_F\,\frac{\alpha_s}{4\pi}\
\frac{1}{\epsilon}\,
\frac{(4\pi)^\epsilon}{\Gamma(1-\epsilon)}\
\left(\frac{\mu^2}{- p^2}\right)^{\!\epsilon}\
\frac{\Gamma(1+\epsilon)\, \Gamma(1-\epsilon)^3\, (1 - \epsilon)}
{\Gamma(1-2\epsilon)\,(1-2\epsilon)}
\eeqn
and where $Z^{(1)}_\psi$ is the one-loop $\overline{\rm MS}$
counter term,
\beqn 
Z^{(1)}_\psi = - C_F\,\frac{\alpha_s}{4\pi}\
\frac{1}{\epsilon}\,
\frac{(4\pi)^\epsilon}{\Gamma(1-\epsilon)}\
\;\;.
\eeqn
Inserting this into Eq.~(\ref{quarkprop1}), we find
\beqn 
i S(p) =  \frac{\mi\s{p}}{p^2 + \mi 0}
\left\{1 + A(p^2) - Z^{(1)}_\psi\right\}
\;\;.
\eeqn

We now want to take $p^2 \to 0$ in order to find $R_\psi$. The
renormalization counter term $Z^{(1)}_\psi$ removes the pole at
$\epsilon = 0$ for any negative value of $p^2$. Having removed the pole,
we can analytically continue $\epsilon$ to ${\rm Re}\, \epsilon < 0$.
Then we have $A(p^2) \to 0$ as $-p^2 \to 0$. This leaves 
\beqn 
i S(p) \sim  \frac{\mi\s{p}}{p^2 + \mi 0}
\left\{1 - Z^{(1)}_\psi\right\}
\;\;.
\eeqn
Comparing to Eq.~(\ref{Rpsidef}), we have
\beqn 
R_\psi^{(1)} = - Z^{(1)}_\psi
=
C_F\,\frac{\alpha_s}{4\pi}\
\frac{1}{\epsilon}\,
\frac{(4\pi)^\epsilon}{\Gamma(1-\epsilon)}\;\;.
\eeqn

The analysis for gluons is similar. One has 
\beqn 
R_A^{(1)} = - Z^{(1)}_A
=
- \left[
\frac{5}{3}\,C_A
- \frac{4}{3}\,T_R\,n_f
\right]
\,\frac{\alpha_s}{4\pi}\
\frac{1}{\epsilon}\,
\frac{(4\pi)^\epsilon}{\Gamma(1-\epsilon)}\
\;\;,
\eeqn
where $T_R = 1/2$ and $n_f$ is the number of quark flavors.

\section{\label{sec:UV}Renormalization}

In the following sections, we will consider the divergences of
infrared origin that can arise in the integral
\beqn
\int\! \frac{d^{d} l}{(2\pi)^{d}}\,
\big| \widetilde{\cal G}(G;\{l\};p_1,\dots,p_m)\big\ra
\eeqn
for a graph $G$. First, however, we should deal with the possible
ultraviolet divergences. These are to be eliminated according to the
$\overline{\rm MS}$ renormalization prescription. However, since we
calculate loop integrals by numerical integration, the implementation of
the $\overline{\rm MS}$ prescription needs some analysis. This is the
subject of the present section.
 
Consider an ultraviolet divergent one-loop graph with $n$ propagators in
the loop. Here $n$ could be 2, 3, or 4. Denote the momenta leaving
$\Gamma$ by $k_1,\dots,k_n$. In our application, $\Gamma$ will be a
subgraph of $G$ and the momenta $k_1,\dots,k_n$ will be linear
combinations of $p_1,\dots,p_m$. The graph $\Gamma$ has the generic form
\beqn
\Gamma(k_1,\dots,k_n) &=&
\int\! \frac{d^{d} l}{(2\pi)^{d}}\,
\widetilde\Gamma(k_1,\dots,k_n;l)
\;\;.
\label{renormstart}
\eeqn
The functions $\Gamma$ and $\widetilde \Gamma$ may carry spinor and vector
indices as well as color indices. Our notation here suppresses these
indices. We seek to calculate the renormalized version of $\Gamma$. With
$\overline{\rm MS}$ renormalization, this is
\beqn
\left[\Gamma\right]_{\rm R}
=
\lim_{\epsilon \to 0}\left\{
\Gamma -
\left[\Gamma\right]_{\rm pole}
\right\}
\;\;,
\label{msbardef1}
\eeqn
where
\beqn
\left[\Gamma\right]_{\rm pole}
 =
\frac{1}{\epsilon}\,\frac{(4\pi)^\epsilon}{\Gamma(1-\epsilon)}\
\times \lim_{\epsilon\to 0}\, [\epsilon\, \Gamma(\epsilon)]
\;\;.
\label{msbardef2}
\eeqn
Here we are to choose the external momenta so that $\Gamma$ does not have
infrared divergences. For instance, all of the external momenta can be
spacelike. Then the only pole present in $\Gamma(\epsilon)$ in
Eq.~(\ref{msbardef2}) is the ultraviolet pole that is to be removed by
renormalization.

Since we are performing integrals numerically, we need to represent 
$\left[\Gamma\right]_{\rm pole}$ as an integral
\beqn
\left[\Gamma\right]_{\rm pole}
 =
\int\! \frac{d^{d} l}{(2\pi)^{d}}\,
\widetilde\Gamma_{UV}(k_1,\dots,k_n;l)
\label{MSbarsubtraction}
\eeqn
in such a way that $\left[\Gamma\right]_R$ can be calculated as
\beqn
\left[\Gamma\right]_R
 =
\int\! \frac{d^{4} l}{(2\pi)^{4}}\,
\lim_{\epsilon \to 0}\left\{
\widetilde\Gamma(k_1,\dots,k_n;l)
-
\widetilde\Gamma_{UV}(k_1,\dots,k_n;l)
\right\}\;\;.
\label{renormcalc}
\eeqn
This last step is justified if the integrands $\widetilde\Gamma$ and
$\widetilde\Gamma_{UV}$ match up to an $l^{-5}$ remainder for $l \to
\infty$ at fixed $\epsilon$ and if $\widetilde\Gamma_{UV}$ if free of
infrared singularities. 

There is more than one possibility for $\widetilde\Gamma_{UV}$. However,
there is a simple prescription that works in all cases save one, the
one-loop gluon self-energy, which has a quadratic divergence and needs a
more elaborate treatment. The prescription, ignoring complications
arising from the tensor structure of $\Gamma$, is as follows. (We give
the full details including the gluon self-energy and the tensor structure
in Appendix \ref{app:renorm}). Write the starting integrand in the form
\beqn
\widetilde\Gamma(k_1,\dots,k_n;l) &=&
\frac{N(k_1,\dots,k_n;l)}
{(l_1^2  + \mi0)\cdots(l_n^2  + \mi0)}
\;\;.
\label{renormintegrand}
\eeqn
Here we display the denominator for each propagator and put everything
else in a numerator function $N(k_1,\dots,k_n;l)$. We make a definite
choice for the loop momentum $l$. In general,
\begin{equation}
l_i = l + K_i\;\;,
\label{lirelation}
\end{equation}
where $K_i$ is a linear combination of the momenta $k_j$ leaving the graph
at the vertices. We define
\begin{equation}
l = \frac{1}{n}[l_1 + l_2 + \cdots + l_n]\;\;.
\label{symmetricchoice}
\end{equation}
Then
\begin{equation}
\sum_i K_i = 0\;\;.
\label{symmetricP}
\end{equation}

The numerator contains an overall factor $\mu^{2\epsilon}$ times a
polynomial in $l$, with coefficients that can be polynomials in
$\epsilon$. Except for the gluon two-point function, which we treat
separately, the highest order term is of order $l^{2n-3}$, corresponding
to a linearly divergent integral for $d = 4$. In some cases, this term is
absent. The next highest order term is of order $l^{2n-4}$, corresponding
to a logarithmically divergent integral. This term is present as long as
we are dealing with a graph that needs renormalization. Call these two
terms
$N_{UV}$,
\begin{equation}
N(k_1,\dots,k_n;l)
= N_{UV}(k_1,\dots,k_n;l) + {\cal O}(l^{2n-5})
\;\;.
\label{NUV}
\end{equation}
Now define
\beqn
\widetilde\Gamma_{UV}(k_1,\dots,k_n;l) &=&
\frac{N_{UV}(k_1,\dots,k_n;l)}
{(l^2  - \mu^2 e^\lambda + \mi0)^n}
\;\;,
\label{theUVsubtraction}
\eeqn
where $\mu$ is the $\overline{\rm MS}$ renormalization scale and
$\lambda$ is a constant to be determined. With this choice,
$\widetilde\Gamma_{UV}$ is free of infrared singularities that could lead
to an infrared divergence. It matches $\widetilde\Gamma$ for large $l$.
Here it is evident that the leading terms, which correspond to the linear
divergence, match. To treat the next terms, note that for $l \to \infty$,
\beqn
\frac{1}{(l + K_1)^2 \cdots (l + K_n)^2} &\sim&
\frac{1}{(l^2)^n}
-
\frac{\sum_i 2 l\cdot K_i}{(l^2)^{n+1}}\,
+
{\cal O}\!\left(
\frac{1}{(l^2)^{n+1}}
\right)
\;\;.
\label{denomexpansion}
\eeqn
The second term vanishes because of Eq.~(\ref{symmetricP}). For this
reason, the order $l^{2n-4}$ term in $\widetilde\Gamma -
\widetilde\Gamma_{UV}$, which would give a logarithmic divergence,
vanishes. Thus $\widetilde\Gamma - \widetilde \Gamma_{UV}$ falls off
fast enough that its integral is ultraviolet finite when $d$ is near
4. Consequently, the $\epsilon \to 0$ limit of the integral of
$\widetilde\Gamma - \widetilde \Gamma_{UV}$ can be evaluated by simply
setting $d = 4$.

We note next that the integral
\beqn
\int\! \frac{d^{d} l}{(2\pi)^{d}}\,
\widetilde\Gamma_{UV}(k_1,\dots,k_n;l)
\eeqn
will be a constant times the  $\overline{\rm MS}$ pole factor
$({1}/{\epsilon})\,{(4\pi)^\epsilon}/{\Gamma(1-\epsilon)}$ plus a
remaining finite piece. The constant multiplying the pole is
necessarily the same constant as for the integral of
$\widetilde\Gamma$ since the ultraviolet behaviors of $\Gamma$ and
$\Gamma_{UV}$ match. The finite constant depends on $\lambda$, so it
is a simple matter to choose $\lambda$ such that the finite part
vanishes, yielding Eq.~(\ref{MSbarsubtraction}).

We have seen here how to construct the renormalization counterterm for a
divergent subgraph $\Gamma$ of a graph $G$. Using this counterterm,
renormalization amounts to replacing
\beqn
\int\! \frac{d^{d} l}{(2\pi)^{d}}\,
\big| \widetilde{\cal G}(G;\{l\};p_1,\dots,p_m)\big\ra
\eeqn
by
\beqn
\int\! \frac{d^{d} l}{(2\pi)^{d}}\,
\left\{
\big| \widetilde{\cal G}(G;\{l\};p_1,\dots,p_m)\big\ra
- \big|\widetilde{\cal R}(G;\{l\};p_1,\dots,p_m)\big\ra
\right\}\;\;,
\label{renormterm}
\eeqn
where $\big|\widetilde{\cal R}(G;\{l\};p_1,\dots,p_m)\big\ra$ is
obtained from $G$ by substituting $\widetilde \Gamma_{UV}$ for
$\widetilde \Gamma$. In the case that graph $G$ is ultraviolet finite,
we define $\big|\widetilde{\cal R}(G;\{l\};p_1,\dots,p_m)\big\ra$ to
be zero. We will use this notation in subsequent sections.

We give specific expressions for $\widetilde\Gamma_{UV}(k_1,\dots,k_n;l)$
in Appendix \ref{app:renorm}. There, we also exhibit what to do with
the tensor structure of the $\widetilde\Gamma(k_1,\dots,k_n;l)$
and with the gluon self-energy diagram, which needs a somewhat different
treatment than presented above because it contains a quadratic divergence.
 
\section{\label{sec:soft}Soft singularities}

The integrand of a one-loop graph becomes singular when the momentum of
an internal gluon loop line that connects to two external lines becomes
soft. Consider a one-loop graph $G$ with external momenta
$p_1,p_2,\dots p_m$ directed out of the graph. Choose two of the external
lines, with labels $i$ and $j$. If the lines $i$ and $j$ connect via
three-point vertices to the two ends of a gluon propagator in the loop,
then we will say that $\{i,j\}$ is in the soft indices class for graph
$G$, $\{i,j\} \in I_S(G)$. In this case the integrand  $\big|
\widetilde{\cal G}(G;\{l\};p_1,\dots,p_m)\big\ra$ is singular in the
limit that  the momentum of the gluon propagator that connects lines $i$
and $j$, call it $l_k$, tends to zero. The loop integration is
logarithmically divergent at this point.

The denominators that become singular in the soft limit $l_k \to 0$ are
$l_k^2$ for the soft gluon propagator, $(l_k+p_i)^2$ for the immediately
preceding propagator in the loop and $(l_k - p_j)^2$ for the immediately
following propagator\footnote{Here we choose the positive direction
  around the loop to be from line $i$ to line $j$.}. It is useful to
define a soft limit function
\beqn
\big|f_{ij}^S (G;p_1,\dots,p_m) \big\ra
=
\lim_{l_k \to 0}
l_k^2 (l_k+p_i)^2(l_k-p_j)^2
\big| \widetilde{\cal G}(G;
l_1,\dots,l_n;p_1,\dots,p_m)\big\ra
\;\;.
\label{softfij}
\eeqn
Here we implicitly take $l_k$ to be the independent loop momentum and use
Eq.~(\ref{loopspace}) to determine the other $l_i$. With this definition,
$\big|f_{ij}^S\big\ra$ vanishes if $\{i,j\} \notin I_S(G)$. We can then
define a soft subtraction for the integrand $\widetilde{\cal G}$ as
\beqn
\big|\widetilde{\cal S}_{ij}(G;l_k;p_1,\dots,p_m)\big\ra 
=
\frac{\big|f_{ij}^S (G;p_1,\dots,p_m)\big\ra}
{(l_k^2 +\mi0)((l_k+p_i)^2 +\mi0)((l_k-p_j)^2 +\mi0)}
\;\;.
\label{softsubtraction}
\eeqn

The integral for the original graph minus this subtraction is
\beqn
\int\! \frac{ d^d l}{(2\pi)^d}\
 \biggl\{
\big| \widetilde{\cal G}(G;\{l\};p_1,\dots,p_m)\big\ra
-
\big|\widetilde{\cal S}_{ij}(G;l_k;p_1,\dots,p_m)\big\ra
\biggr\}
\;\;.\ \
\eeqn
By construction, this integral does {\it not} have an infrared divergence
from the region $l_k \to 0$. It may have other divergences, which will be
eliminated by other subtraction terms. For instance, there may be another
pair of external lines $\{i',j'\}$ that connect to the virtual loop and
are joined by a virtual gluon line with label $k'$. Then we should also
subtract $\big|{\cal S}_{i'j'}\big\ra$, as defined above, from the graph.
There may also be collinear divergences, to be discussed in the following
section. Finally, there may be ultraviolet divergences, as discussed in
Sec.~\ref{sec:UV}. The idea is that after all divergences are subtracted,
the indicated loop integral can be performed by numerical integration.

Having subtracted the integral of $\big|\widetilde{\cal S}_{ij}\big\ra$,
we should add it back. For this purpose, we need to study the structure of
the soft subtraction in more detail. The first step is to extend the
vector notation that we have been using to cover unphysical gluon
polarizations. Previously we have implicitly supplied a projection
operator $\big|s\big\ra \big\la s\big|$ onto the $d - 2$ dimensional space
of physical polarizations for each external gluon. For the purpose of the
present section we consider a $d$-dimensional polarization space.

In addition to the physical polarizations represented by $\big|s\big\ra$
we introduce polarizations $\big|\mu\big\ra$ along the coordinate axes.
These have the orthogonality relations
\beqn
\big\la \mu
\big| \nu\big\ra = - g_{\mu\nu}
\;\;.
\label{orthoganality}
\eeqn
The minus sign is present because we have chosen the inner product in the
bra-ket spin space so that $\big\la s' \big| s\big\ra = + \delta_{s's}$.
The corresponding completeness relation for the $\big| \mu\big\ra$
vectors is
\beqn
1 =
-\big|\mu\big\ra
g^{\mu\nu}
\big\la\nu\big| 
\label{completeness}
\eeqn
with an implicit summation. We take the sign convention for the vectors 
$\big| \mu\big\ra$ such that
\beqn
\big\la\mu\big| {\cal M}\big\ra = M_\mu
\;\;.
\label{mubradef}
\eeqn
Then our previous definition $\big\la s\big| {\cal M}\big\ra =
\varepsilon^*(p,s)^\mu M_\mu$ can be written
\beqn
\big\la s \big| {\cal M}\big\ra = 
\varepsilon^*(p,s)^\mu
\big\la \mu \big| {\cal M}\big\ra
\;\;.
\eeqn

The soft limit function has a rather simple form
\beqn\label{new:softfij}
\big|f_{ij}^S (G;p_1,\dots,p_m)\big\ra = 
-\mi \gs^2\mu^{2\epsilon} 4p_i\!\cdot\! p_j
\bom{T}_i\cdot\bom{T}_j \,\bmi{H}_{ij}
\big| {\cal G}(G_{ij}(G);p_1,\dots,p_m)\big\ra
\;\;,\quad\;
\eeqn
where the graph $G_{ij}(G)$ represents the tree level amputated graph
obtained from graph $G$ by omitting the three singular propagators and the
vertices where they join the external lines $i$ and $j$. Here
$\bmi{H}_{ij}$ is  an operator in the spin space of the partons
$i$ and $j$. The form of $\bmi{H}_{ij}$ depends on whether partons $i$ and
$j$ are quarks (or equivalently antiquarks) or gluons. In computing
$\bmi{H}_{ij}$, we keep only terms that survive under projecting onto
physical polarizations for the final state partons. We find
\beqn\label{Hop:qq}
\bmi{H}_{\!ij}(qq)  &=& 1
\;\;,\nonumber\\
\bmi{H}_{\!i j}(qg)  &=& 
1 +
\big| \nu \big\ra_{\!j}\,
\frac{ p_{i}^{\nu}\ p_{j}^{\beta}}
{2 p_i\cdot p_j}\
\big\la_{\!\!\!\!\!j}\,\beta\big|
\;\;,\nonumber\\
\bmi{H}_{\!ij}(gg)  &=& 
1
+
\big| \mu \big\ra_{\!i}\,
\frac{ p_{j}^{\mu}\ p_{i}^{\alpha}}
{2 p_i\cdot p_j}\
\big\la_{\!\!\!\!\!i}\,\alpha\big|
+
\big| \nu \big\ra_{\!j}\,
\frac{ p_{i}^{\nu}\ p_{j}^{\beta}}
{2 p_i\cdot p_j}\
\big\la_{\!\!\!\!\!j}\,\beta\big|
+
\big| \mu \big\ra_{\!i}\big| \nu \big\ra_{\!j}
\frac{\,g^{\mu\nu}\
p_{i}^{\alpha}\,p_{j}^{\beta}\ }{4 p_i\cdot p_j}\
\big\la_{\!\!\!\!\!i}\,\alpha\big|\,
\big\la_{\!\!\!\!\!j}\,\beta\big|
\;.
\label{Hop}
\eeqn
Notice that in each term beyond the unit operator there is a factor
$p_{j}^{\beta}\ \, \big\la_{\!\!\!\!\!j}\,\beta\big|$ or $p_{i}^{\alpha}\
\,\big\la_{\!\!\!\!\!i}\,\alpha\big|$. This feature will play an important
role in the subsequent analysis.

With this result, the soft subtraction is
\beqn
\big|\widetilde{\cal S}_{ij}(G;l_k;p_1,\dots,p_m)\big\ra 
=
E_{ij}(l_k,p_i,p_j)\,\bom{T}_i\cdot\bom{T}_j\,
\bmi{H}_{ij}\,
\big| {\cal G}(G_{ij}(G);p_1,\dots,p_m)\big\ra
\;\;,
\label{softsubtractionquark}
\eeqn
where the $E_{ij}(l_k,p_i,p_j)$ is the familiar eikonal factor
\cite{Grammer:1973db,Giele:1992vf},
\beqn\label{soft:eikonal}
E_{ij}(l_k,p_i,p_j) 
= -\mi \gs^2\mu^{2\epsilon} \frac{4p_i\cdot p_j}
{(l_k^2 +\mi0)((l_k+p_i)^2 +\mi0)((l_k-p_j)^2 +\mi0)}
\;\;.
\eeqn
Note that, in $E_{ij}$, one could arrange the definitions so that the
denominator $(l_k+p_i)^2 +\mi0 = 2 l_k\cdot p_i + l_k^2 +\mi0$ is replaced
by $2 l_k\cdot p_i +\mi0$,  as in \cite{Grammer:1973db}, since these
expressions match in the limit $l_k \to 0$. Similarly $(l_k - p_j)^2
+\mi0$ could have been $-2 l_k\cdot p_j +\mi0$. The form used in
Eq.~(\ref{soft:eikonal}) suits our purpose here because its integral over
$l_k$ is ultraviolet convergent. The subtraction term in
Eq.~(\ref{softsubtractionquark}) is illustrated in Fig.~\ref{fig:soft}.

\FIGURE[ht]{
$\dd\lim_{l_k\to 0}\frac{1}{E_{ij}(l_k)}\;
\begin{array}{c}
  \begin{picture}(80,80)(-20,-40)
   \LongArrowArc(15,2)(17,-45,45)
    \Vertex(40,20){1}
    \Vertex(40,-20){1}
    \ArrowLine(15,25)(40,20)
    \ArrowLine(40,20)(60,35)
    \Gluon(15,-25)(40,-20){4}{4}
    \Gluon(40,20)(40,-20){4}{6}
    \Gluon(40,-20)(60,-35){4}{4}
    \GOval(0,0)(40,20)(0){0.9}
    \Text(0,0)[]{$G_{ij}$}
    \Text(62,35)[l]{$p_i$}
    \Text(62,-35)[l]{$p_j$}
    \Text(48,0)[l]{$l_k$}
  \end{picture}
\end{array}  \qquad=\quad
\begin{array}{c}
  \begin{picture}(80,80)(-20,-40)
    \ArrowLine(15,25)(50,35)
    \Gluon(15,-25)(50,-35){4}{6}
    \GOval(0,0)(40,20)(0){0.9}
    \DashCArc(0,0)(42,-38,44){3}
    \DashCArc(0,0)(40,-40,46){3}
    \Text(0,0)[]{$G_{ij}$}
    \Text(52,35)[l]{$p_i$}
    \Text(52,-35)[l]{$p_j$}
  \end{picture}
\end{array}\quad - \quad
\dd \frac{p_i\cdot\varepsilon(p_j)}{2p_i\cdot p_j}\;
\begin{array}{c}
  \begin{picture}(80,80)(-20,-40)
    \ArrowLine(15,25)(50,35)
    \ArrowGluonn(16,-25)(50,-35){4}{4}
    \GOval(0,0)(40,20)(0){0.9}
    \DashCArc(0,0)(40,-40,46){3}
    \DashCArc(0,0)(42,-38,44){3}
    \Text(0,0)[]{$G_{ij}$}
    \Text(52,35)[l]{$p_i$}
    \Text(52,-35)[l]{$p_j$}
  \end{picture}
\end{array}$
\caption{\label{fig:soft}Pictorial representation of the soft gluon
 subtraction, Eq.~(\ref{softsubtractionquark}). The quantity depicted is
$(1/E_{ij})\big|{\cal S}_{ij}\big\ra$ for $i = q$, $j = g$. The double
dashed line represents the octet color exchange contained in the factor
$\bom{T}_i\cdot\bom{T}_j$. The two terms correspond to the two terms in
$\bmi{H}_{ij}(qg)$. In the second term, the arrow on the gluon line
represents the factor  $p_{j}^{\beta}\ \,
\big\la_{\!\!\!\!\!j}\,\beta\big|$, which has the effect of replacing the
polarization vector $\epsilon^*(p_j,s)$ that would normally be contracted
with the amputated Green function by the momentum $p_j$. The polarization
vector is then contracted with $p_i$. The minus sign arises from our sign
conventions, in which $\big\la s
\big|\nu\big\ra_{\!j}\ p_i^\nu = - \epsilon^*(p_j,s)\cdot p_i$} }

We need to add back the sum over graphs of the integral $\big|{\cal
S}_{ij}\big\ra$ of $\big|\widetilde{\cal S}_{ij}\big\ra$. We are now 
prepared to calculate what this quantity is. Defining ${\cal
V}_{ij}^{\mathrm{soft}}(\epsilon)$ to be the integral of $E_{ij}(l)$, we
have
\beqn
\big|{\cal S}_{ij}(G;p_1,\dots,p_m)\big\ra 
&\equiv& 
\int \frac{d^dl}{(2\pi)^d}\
\big|\widetilde{\cal S}_{ij}(G;\{l\},p_1,\dots,p_m)\big\ra
\nonumber \\
&=&
{\cal V}_{ij}^{\mathrm{soft}}(\epsilon) \,
\bom{T}_i\cdot\bom{T}_j\,
\bmi{H}_{ij}\,
\big| \widetilde{\cal G}(G_{ij}(G);p_1,\dots,p_m)\big\ra
\;\;.
\eeqn
The integral can be performed analytically, with the result
\beqn
{\cal V}_{ij}^{\mathrm{soft}}(\epsilon) 
\equiv 
\int\frac{d^dl}{(2\pi)^d}\ E_{ij}(l) =  \frac{\as}{4\pi}
\frac{(4\pi)^\epsilon}{\Gamma(1-\epsilon)}
\left(\frac{\mu^2}{- 2 p_i\!\cdot\! p_j}\right)^{\!\epsilon}
\left(\frac2{\epsilon^2} +{\cal O}(\epsilon)\right)\;\;.
\eeqn
Thus $\big|{\cal S}_{ij}(G)\big\ra$ is a known integral times a tree level
graph with a modified color factor and a kinematic factor
$\bmi{H}_{ij}$. 

The sum over graphs $G$ of the subtraction terms is even simpler. We
consider a given set of final state parton flavors and a given choice of
$\{i,j\}$. The tree graphs $G_{ij}(G)$ have the same parton flavors and
momenta. Summing over all graphs $G$ for which $\{i,j\} \in I_S(G)$, the
graphs $G_{ij}(G)$ cover all tree graphs with this choice of external
partons. Thus summing over graphs $G$ gives the full tree amplitude
$\big|{\cal M}_{\rm tree}(p_1,\dots,p_m)\big\ra$ corresponding to the
given final state. On the other hand the tree level matrix element is
gauge invariant, that is
\beqn
p_{j}^{\mu}\ \big\la_{\!\!\!\!\! j}\ \mu 
\big |{\cal M}_{\rm tree}(p_1,\dots,p_m)\big\ra = 0\;\;,
\eeqn
where $j$ is a gluon leg. Thus gauge invariance ensures that the
complicated terms in the operator $\bmi{H}_{ij}$, Eq.~(\ref{Hop}), do not
contribute, so that $\bmi{H}_{ij}$ can be replaced by the unit operator.
Then the soft contribution that we need to add back is 
\beqn
\sum_G\big|{\cal S}_{ij}(G;p_1,\dots,p_m)\big\ra 
=
{\cal V}_{ij}^{\mathrm{soft}}(\epsilon) \,
\bom{T}_i\cdot\bom{T}_j\,
\big| {\cal M}_{\rm tree}(p_1,\dots,p_m)\big\ra
\;\;.
\label{softcorrections}
\eeqn

We now summarize the soft gluon subtractions, this time including a sum
over pairs of indices $\{i,j\}$. We subtract
\beqn
\sum_{\{i,j\} \in I_S(G)}
\big|\widetilde{\cal S}_{ij}(G;l_k;p_1,\dots,p_m)\big\ra
\;\;,
\label{softsubtractionsum}
\eeqn
defined in Eq.~(\ref{softsubtraction}) from the integrand $\big|
\widetilde{\cal G} (G;\{l\}, p_1,\dots,p_m)\big\ra$ for each one-loop
graph $G$. Then we add the integrals of these terms back in the form
\beqn
\sum_{\{i,j\} }
{\cal V}_{ij}^{\mathrm{soft}}(\epsilon) \,
\bom{T}_i\cdot\bom{T}_j\,
\big| {\cal M}_{\rm tree}(p_1,\dots,p_m)\big\ra
\;\;.
\label{softadbacksum}
\eeqn
In Sec.~\ref{sec:final}, we will see that the terms added back are
cancelled by similar terms corresponding to real soft gluon emission. The
terms subtracted from each one-loop graph serve to remove the soft gluon
divergences. This leaves the collinear parton divergences, which are the
subject of the next section.

\section{\label{sec:collinear}Collinear singularities}

One-loop graphs have logarithmic infrared divergences that arise from
integration regions in which  the momentum on an internal loop line that
connects to an external line becomes collinear with the momentum of the
external line. In this section, we define a term that, when subtracted
from the graph, eliminates the divergence. This construction is quite
similar to what we needed for soft gluon divergences. However, the
structure of the collinear subtraction term is more complicated, so we
will need a more involved analysis. 

Consider a one-loop graph $G$ with external momenta $p_1,p_2,\dots p_m$
directed out of the graph. Choose one of the external lines, with label
$i$. If the line $i$ connects via a three-point vertex to the loop, and
if the loop partons are both gluons or are one gluon and one quark or
antiquark, then we will say that $i$ is in the collinear index class for
graph $G$, $i \in I_C(G)$. In this case the loop integration has a
logarithmic divergence arising from the region in which the momenta of
the loop propagators that connect to line $i$, call them $j$ and $j+1$,
become collinear to the outgoing momentum of line $i$,
\beqn
l_j &\to& x\, p_i
\nonumber\\
-l_{j+1} &\to& (1-x)\, p_i
\;\;,
\label{collinearlimit}
\eeqn
with $0<x<1$.

The denominators that become singular in the collinear limit
(\ref{collinearlimit}) are $l_j^2$ and $(-l_{j+1})^2 = (p_i - l_j)^2$. The
coefficient of $1/[l_j^2\times(p_i - l_j)^2]$ in the integrand for the
graph is non-singular in the collinear limit. It is useful to define a
collinear coefficient function $\big| f^{C,0}_i(G;x;p_1,\dots,p_m)\big\ra$
by
\beqn
\big| f^{C,0}_i(G;x;p_1,\dots,p_m)\big\ra
= 
\lim_{l_j \to x\, p_i}
l_j^2\,(p_i - l_j)^2\,
\big| \widetilde{\cal G}(G;l_1,\dots,l_{n};p_1,\dots,p_m)\big\ra
\;\;.
\label{collinearcoefficient1}
\eeqn
In writing Eq.~(\ref{collinearcoefficient}), we implicitly take $l_j$ to
be the independent loop momentum and use Eq.~(\ref{loopspace}) to
determine the other $l_i$ in $\big| \widetilde{\cal
G}(G;l_1,\dots,l_{n};p_1,\dots,p_m)\big\ra$. In particular, $l_{j+1} = l_j
- p_i$.

One can use the collinear coefficient function to remove the collinear
divergence. Consider the integral
\beqn
&&\int\! \frac{ d^d l}{(2\pi)^d}\
\biggl\{
\big| \widetilde{\cal G}(G;\{l\},p_1,\dots,p_m)\big\ra
\nonumber\\
&&
- \frac{1}{(l_j^2  + \mi0)((p_i - l_j)^2 + \mi0)}\
\int_0^1\!dx\ \delta\!\left(x - \frac{l_j\cdot n_i}{p_i\cdot n_i}\right)
\big| f^{C,0}_i(G;x;p_1,\dots,p_m)\big\ra
\biggr\}\;\;. 
\label{naivecollinear}
\eeqn
Here we define the momentum fraction $x$ away from the collinear limit by
using a lightlike vector $n_i$. A good choice for $n_i$ is
\beqn
n_i^\mu = -p_i^\mu + \frac{2\,p_i\!\cdot\!w}{w^2} w^\mu  \;\;,\qquad
w^\mu = \sum_{k\in {\rm final\ state}} p_k^\mu\;\;. 
\label{nvecdef}
\eeqn
By construction, the integral in Eq.~(\ref{naivecollinear})
does {\it not} have a divergence from the collinear region
(\ref{collinearlimit}). The divergence was only logarithmic and the
subtraction removes the leading singularity, leaving at worst an
integrable singularity. 

There are two problems with  Eq.~(\ref{naivecollinear}).  The first is
that we have already subtracted something from the integrand for the graph
$G$, so we do not start with a clean slate. In the present notation, the
soft subtraction term corresponding to propagator $j$ in the loop being
soft is, in the collinear limit (\ref{collinearlimit}),
\beqn
\frac{1}{(l_j^2  + \mi0)((p_i - l_j)^2 + \mi0)}\
\frac{1}{x}\,\lim_{y \to 0}\, y
\big| f^{C,0}_i(G;y;p_1,\dots,p_m)\big\ra
\;\;.
\label{collinearsoft1}
\eeqn
(If the parton in loop propagator $j$ is not a gluon, then this quantity
is zero.) Similarly the soft subtraction term corresponding to propagator
$j+1$ in the loop being soft is, in the collinear limit
(\ref{collinearlimit}),
\beqn
\frac{1}{(l_j^2  + \mi0)((p_i - l_j)^2 + \mi0)}\
\frac{1}{1-x}\,\lim_{y \to 1}\, (1-y)
\big| f^{C,0}_i(G;y;p_1,\dots,p_m)\big\ra
\;\;.
\label{collinearsoft2}
\eeqn
In order to cancel these contributions to the net integrand in the
collinear limit, we should subtract them from the collinear subtraction
term in Eq.~(\ref{naivecollinear}). To this end, we define a revised
collinear coefficient function $\big| f^C_i(G;x;p_1,\dots,p_m)\big\ra$
\beqn
&&\big| f^C_i(G;x;p_1,\dots,p_m)\big\ra
= 
\big| f^{C,0}_i(G;x;p_1,\dots,p_m)\big\ra
\nonumber\\
&&\ \ \ \ \ \ \ \ \
-\frac{1}{x}\, \lim_{y\to 0}\ y
\big| f^{C,0}_i(G;y;p_1,\dots,p_m)\big\ra
-\frac{1}{1 - x}\, \lim_{y\to 1}\ (1-y)
\big| f^{C,0}_i(G;y;p_1,\dots,p_m)\big\ra
\;.\ \ \ \ \ 
\label{collinearcoefficient}
\eeqn
With this coefficient function, our subtraction inside the loop integral
is
\beqn
\frac{1}{(l_j^2  + \mi0)((p_i - l_j)^2 + \mi0)}
\int_0^1\!dx\ \delta\!\left(x - \frac{l_j\cdot n_i}{p_i\cdot n_i}\right)
\big| f^C_i(G;x;p_1,\dots,p_m)\big\ra
\;\;.
\label{naivecollinear2}
\eeqn

A second problem remains with Eq.~(\ref{naivecollinear2}). The integral
of the subtraction term is ultraviolet divergent. We can easily fix that
by modifying the subtraction term to be
\beqn
\frac{f_{UV}(l_j, l_j - p_i)}{(l_j^2  + \mi0)((p_i - l_j)^2 + \mi0)}
\int_0^1\!dx\ \delta\!\left(x - \frac{l_j\cdot n_i}{p_i\cdot n_i}\right)
\big| f^C_i(G;x;p_1,\dots,p_m)\big\ra
\;\;,
\label{naivecollinear3}
\eeqn
where
\beqn
f_{UV}(l_j, l_j - p_i) 
= \frac{1}{2} \left(\frac{-\mu^2e}{l_j^2-\mu^2e+\mi0}+
\frac{-\mu^2e}{(l_j - p_i)^2-\mu^2e+\mi0}\right)\;\;.
\eeqn
Here $\mu$ is the $\overline{\rm{MS}}$ renormalization scale and
$e=2.71828\dots$ is the base of natural logarithms. The factor
$f_{UV}$ provides an extra power of $l_j^2$ in the denominator for large
$l_j$ but equals 1 in the collinear limit.

In summary, we subtract from the integrand for each graph $G$ collinear
subtraction terms
\beqn
\sum_{i\in I_C(G)}
\big| \widetilde{\cal C}_i(G;\{l\},p_1,\dots,p_m)\big\ra\;\;,
\eeqn
where
\beqn
\big| \widetilde{\cal C}_i(G;\{l\},p_1,\dots,p_m)\big\ra
&=&
\frac{f_{UV}(l_j, l_j - p_i)}{(l_j^2  + \mi0)((p_i - l_j)^2 + \mi0)}
\nonumber\\
&&\times
\int_0^1\!dx\ \delta\!\left(x - \frac{l_j\cdot n_i}{p_i\cdot n_i}\right)
\big| f^C_i(G;x;p_1,\dots,p_m)\big\ra
\;\;.
\label{collinearsubtraction}
\eeqn
These subtractions, together with the soft subtractions, remove all of
the infrared divergences from the loop integrals. One should note that,
although the definition, Eq.~(\ref{collinearcoefficient}), of the
collinear subtraction terms seems rather complicated, it is a purely
algebraic construction that can be accomplished by straightforward
computer algebra.

Our next task, pursued in the following two subsections, will be to add
back the subtraction terms $\big| \widetilde{\cal
C}_i(G;\{l\},p_1,\dots,p_m)\big\ra$, this time integrated and summed over
graphs $G$. That is, we want to add back the sum over graphs of
\beqn
\big| {\cal C}_i(G;\{p_1,\dots,p_m)\big\ra
\equiv\int\frac{d^dl}{(2\pi)^d}\
\big| \widetilde{\cal C}_i(G;\{l\},p_1,\dots,p_m)\big\ra
\;\;.
\label{integratedCdef}
\eeqn

\subsection{\label{sec:collinearquark}Quark line}

For each one-loop virtual graph $G$ and for each of its external lines
$i$ with $i\in I_C(G)$, we have subtracted a quantity $\big| {\cal
C}_i(G;p_1,\dots,p_m)\big\ra$ from $\big|{\cal G} (G;p_1,\dots,p_m)
\big\ra$. These subtractions soften the collinear divergences in
$\big|{\cal G}(G;p_1,\dots,p_m)\big\ra$ with its soft gluon subtractions,
so that one could perform the loop integral for graph $G$ by numerical
integration. Now we must add the collinear subtractions $\big| {\cal
C}_i(G;p_1,\dots,p_m)\big\ra$ back, this time performing the integral
analytically. By design, these integrals have collinear singularities
that give poles at $d=4$ dimensions. The idea is to cancel these poles
against the poles from the subtraction terms for collinear divergences in
NLO tree graphs.

Unfortunately, the structure of the $\big| {\cal C}_i (G;p_1,\dots,p_m)
\big\ra$ for any given graph $G$ is quite complicated. Fortunately, the
structure of the sum over graphs $G$ of $\big| {\cal
C}_i(G;p_1,\dots,p_m)\big\ra$ is vastly simpler. One gets a singular
factor times the leading order matrix element $\big|{\cal M}
(p_1,\dots,p_m)\big\ra$ calculated from tree graphs. To see
this, we simply need to apply the construction used to prove
factorization of collinear parton contributions in hadron-hadron
collisions \cite{Collins:1985ue}.

In this subsection, we consider the case that line $i$ is a quark line.
The case that $i$ is an antiquark line follows trivially and is covered
at the end of this section. The case that $i$ is a gluon line is treated
in the following subsection.

\FIGURE[ht]{
$\dd
\lim_{l_j\to xp_i} l_j^2 (l_j-p_i)^2 
\begin{array}{c}
  \begin{picture}(100,80)(-20,-40)
   \LongArrowArc(15,2)(17,-45,45)
    \Vertex(50,0){1}
    \ArrowLine(13.8,29)(50,0)
    \ArrowLine(50,0)(80,0)
    \Gluon(13.8,-29)(50,0){4}{7}
    \GOval(0,0)(40,20)(0){0.9}
    \Text(30,-25)[lt]{\small $l_j$}  
    \Text(28,20)[lb]{\small $l_{j}-p_i$}
   \Text(80,5)[b]{$p_i$}  
   \Text(80,-5)[t]{$\alpha$}  
   \Text(0,0)[]{$\widetilde{\bom{V}}(G_i)$}
  \end{picture}
\end{array} \;\; = \;\;
\dd 2\mi\gs\mu^\epsilon\,T_{\alpha\alpha'}^a
\frac{\sqrt{1-x}}x\;
\begin{array}{c}
  \begin{picture}(100,80)(-20,-40)
    \ArrowLine(17,20)(60,30)
    \ArrowGluonn(18,-20)(60,-30){4}{6}
    \GOval(0,0)(40,20)(0){0.9}
    \Text(60,30)[bc]{$(1-x)p_i$}  
    \Text(60,28)[tc]{$\alpha'$}  
    \Text(58,-25)[lb]{$xp_i$}
    \Text(58,-35)[lt]{$a$}
    \Text(0,0)[]{$\widetilde{\bom{V}}(G_i)$}
  \end{picture}
\end{array} 
$
\caption{\label{fig:collquark}Illustration of
Eq.~(\ref{collinearcoefficient1Q}). The left hand side represents
$f^{C,0}_i(G;x;p_1,\dots,p_m)^\alpha$, which is defined in
Eq.~(\ref{collinearcoefficient1}) as the collinear limit of
$l_j^2\,(l_{j}-p_i)^2$ times the integrand $\widetilde{\cal
G}(G;l_1,\dots,l_{n};p_1,\dots,p_m)^\alpha$. Here the integrand 
$\widetilde{\cal G}$ consists of a spinor
$\bar u(p_i)$, a vertex and two propagators times
an amputated Green function $\widetilde{\bom{V}}$, as in
Eq.~(\ref{collinearquark0}). The illustration represents the integrand: an
integration over the loop momentum is {\it not} implied.  On the right
hand side, the gluon line of $\widetilde{\bom{V}}$ is contracted with $(x
p_i)_\mu$, indicated by the arrowhead, while the quark line is contracted
with $\bar u((1-x)p_i)$. The remaining factors in the right hand side of
Eq.~(\ref{collinearcoefficient1Q}) are indicated.} }

We begin by setting up a useful notation. By assumption, line $i$,
carrying momentum $p_i$ out of the graph, connects to a virtual loop in
$G$. Line $i$ must therefore connect to the loop at a quark-gluon-quark
vertex. We choose to label the propagators in the loop so that the gluon
line in the loop has label $j$ and the quark line has label $j+1$. Thus
a gluon line with label $j$ carries momentum $l_j$ into the vertex and the
quark line with label $j+1$ carries momentum $-l_{j+1} = p_i - l_j$ into
the vertex. We write the integral for the graph as
\beqn \nn
{\cal G}(G,p_1,\dots,p_m)^\alpha &=&\int\!\frac{d^dl}{(2\pi)^d}\
\widetilde{\cal G}(G;l_1,\dots,l_n,p_1,\dots,p_m)^\alpha
\\
&=& \mi\gs\mu^{\epsilon}
T^a_{\alpha\alpha'}
\int\! \frac{d^dl_j}{(2\pi)^d}\
\frac{1}
{(l_j^2  + \mi0)((p_i - l_j)^2 + \mi0)}
\nonumber \\ &&\qquad\times
\bar{u}(p_i)\gamma_\mu(\s{p}_i - \s{l}_j)
\big[
\widetilde{\bom{V}}^\mu\!(G_i,p_1,\dots,p_i-l_j,\dots,p_m,l_j)
\big]_{\alpha'}^a
\;\;.\qquad
\label{collinearquark0}
\eeqn
Here we have abandoned our vector notation for spin and color space and
written the spin and color indices that are needed for the calculation
explicitly. We display a quark color index $\alpha$ for quark line
$i$ in ${\cal G}$, leaving the other indices on ${\cal G}$ unwritten. The
amplitude $\widetilde{\bom{V}}$ has a color index $a$ corresponding to
the gluon $j$ in the loop and a quark color index
$\alpha'$ corresponding to the quark line $j+1$ in the loop. There is an
explicit color matrix $T^a_{\alpha\alpha'}$ connecting the colors of 
$\widetilde{\bom{V}}$ to the color of ${\cal G}$. The amplitude
$\widetilde{\bom{V}}$ carries a vector index $\mu$ corresponding to the
polarization of the gluon line $j$. It also carries a Dirac spinor index
for the quark line $j+1$. However, we use the matrix notation for the
Dirac structure of this line without displaying the Dirac indices
explicitly. The Dirac spinor $\bar{u}(p_i)$ represents the final state
quark on line $i$. We have displayed the integration over the loop
momentum $l_j$, the quark-gluon-quark vertex and the propagators for the
quark and the gluon in the loop. Everything else is included in the
Feynman amplitude $\widetilde{\bom{V}}$ for the amputated tree level 
graph $G_i$ obtained by omitting the propagators $j$ and $j+1$ and the
vertex that attaches these propagators to external line $i$ in graph
$G$.

Thus the amplitude $\widetilde{\bom{V}}$ is defined by
Eq.~(\ref{collinearglue0}). The momentum arguments of
$\widetilde{\bom{V}}$ are the external momenta of the original graph,
plus the momentum $p_i-l_j$ of the the quark line and the momentum $l_j$
of the gluon line.

With the notation thus defined, we are ready to calculate. The
subtraction term $f^{C,0}$ defined in Eq.~(\ref{collinearcoefficient1})
can be expressed in terms of $\widetilde{\bom{V}}$ as
\beqn
&&f^{C,0}_i(G;x;p_1,\dots,p_m)^\alpha =
\lim_{l_j \to x\, p_i}
l_j^2\,(p_i - l_j)^2\,
\widetilde{\cal G}(G;l_1,\dots,l_{n};p_1,\dots,p_m)^\alpha
\nonumber\\
&&\qquad\qquad=
\mi\gs\mu^{\epsilon}
T^a_{\alpha\alpha'}
\bar{u}(p_i)\gamma_\mu(1-x)\s{p}_i
\,\big[
\widetilde{\bom{V}}^\mu\!(G_i,p_1,\dots,(1-x)p_i,\dots,p_m,xp_i)
\big]_{\alpha'}^a
\nonumber\\
&&\qquad\qquad=
2\mi\gs\mu^{\epsilon}
T^a_{\alpha\alpha'}
\frac{\sqrt{1-x}}{x}\
(x p_i)_\mu\,
\bar{u}((1-x)p_i)
\nonumber\\
&&\qquad\qquad\qquad\qquad\qquad\times
\big[
\widetilde{\bom{V}}^\mu\!(G_i,p_1,\dots,(1-x)p_i,\dots,p_m,xp_i)
\big]_{\alpha'}^a
\;\;.
\label{collinearcoefficient1Q}
\eeqn
In the last step, we rewrite $\bar u(p_i)$ as $\bar
u((1-x)p_i)/\sqrt{1-x}$ in order to have a spinor evaluated at the
momentum of the corresponding quark line. This equation is illustrated in
Fig.~\ref{fig:collquark}.

We would now like to sum over graphs. We  consider one-loop graphs $G$
with $m$ external particles having flavors $\{f_1,\dots,f_m\}$ and
momenta $\{p_1,\dots,p_m\}$. We are considering the collinear subtraction
for parton $i$, which we assume here is a quark. Our graphs are amputated
on their external legs. Furthermore, we need consider only graphs that
have a collinear divergence for external leg $i$: $i\in I_C(G)$. Let us
call this class of graphs $C$.

The corresponding graphs $G_i$ are amputated tree graphs with $m+1$
external legs. The flavors of the first $m$ external particles are
$\{f_1,\dots,f_m\}$, the same as for the graph $G$. The momenta of
these particles are the same as for $G$ except for particle $i$, which
carries momentum $(1-x)p_i$ instead of $p_i$. The external particle with
index $m+1$ is a gluon with momentum $xp_i$. Let us call the set of all
such graphs $C'$. When we sum over all graphs $G \in C$, the graphs
$G_i$ include all graphs in $C'$ except for the graphs in which gluon
$m+1$ couples directly to quark $i$. These graphs are absent
because graphs with a self-energy insertion on external line $i$ are {\it
not} included in $C$. Let us call the set of ($m+1$)-particle graphs
that we do get $C'_+$ and the graphs that we don't get $C'_-$. 

The collinear subtraction defined in Eq.~(\ref{collinearsubtraction}),
summed over graphs $G \in C$ is 
\beqn
\lefteqn{\sum_{G\in C} {\cal C}_i(G;p_1,\dots,p_m)^\alpha}
\nonumber\\&=&
\int\! \frac{ d^d l_j}{(2\pi)^d}\
\frac{f_{UV}(l_j, l_j - p_i)}{(l_j^2  + \mi0)((p_i - l_j)^2 + \mi0)}
\int_0^1\!dx\ \delta\!\left(x - \frac{l_j\cdot n_i}{p_i\cdot n_i}\right) 
\sum_{G \in C}
f^C(G;x;p_1,\dots,p_m)^\alpha
\nonumber\\
&=&
\int\! \frac{ d^d l_j}{(2\pi)^d}\
\frac{f_{UV}(l_j, l_j - p_i)}{(l_j^2  + \mi0)((p_i - l_j)^2 + \mi0)}
\int_0^1\!\frac{dx}{x}
\ \delta\!\left(x - \frac{l_j\cdot n_i}{p_i\cdot n_i}\right)
2\mi\gs\mu^{\epsilon}\,
T^a_{\alpha\alpha'} 
\nonumber\\
&&\qquad\qquad\times
\biggl\{
(1-x)\,H_{\alpha'}^a(x;p_1,\dots,p_m)
-
H_{\alpha'}^a(0,p_1,\dots,p_m)
\biggr\}
\;\;.
\label{collinearsubtractionfromH}
\eeqn
Here we define
\beqn
&&H_{\alpha'}^a(x;p_1,\dots,p_m)
=
\nonumber\\
&&\hskip 0.5 cm
(1-x)^{-1/2}\
\lim_{q \to x p_i}
\sum_{G_i \in C'_+}
q_\mu\,
\bar{u}((1-x) p_i)
\big[
\widetilde{\bom{V}}^\mu\!(G_i,p_1,\dots,(1-x)p_i,\dots,p_m,q)
\big]_{\alpha'}^a
\;\;.\quad\quad
\label{Hqdef}
\eeqn
Note that we have written the momentum of parton $m+1$ as $q$ instead of
$xp_i$. Then we take the limit as $q \to xp_i$. This avoids an ambiguous
expression of the form $0/0$ in the subsequent calculation.

\FIGURE[ht]{
$\dd\frac{-1}{\dd \sqrt{1-x}}
\sum_{G_i\in C'_-} \lim_{q\to xp_i}\;
\begin{array}{c}
  \begin{picture}(100,90)(-20,-45)
    \ArrowLine(17,20)(60,30)
    \ArrowGluonn(18,-20)(60,-30){4}{6}
    \GOval(0,0)(40,20)(0){0.9}
    \Text(60,30)[bc]{$(1-x)p_i$}  
    \Text(60,28)[tc]{$\alpha'$}  
    \Text(58,-25)[lb]{$q$}
    \Text(58,-35)[lt]{$a$}
    \Text(0,0)[]{$\widetilde{\bom{V}}(G_i)$}
  \end{picture}
\end{array} 
$
\hfill~\\
~ 
\hfill 
$~\qquad\qquad\qquad =\dd \frac{-1}{\dd \sqrt{1-x}} 
\lim_{q\to xp_i} 
\sum_{G'_i\in C_0}\;
\begin{array}{c}
  \begin{picture}(100,90)(-20,-45)
    \ArrowLine(17,20)(60,30)
    \ArrowGluonn(31,22.6)(60,-30){4}{8}
    \GOval(0,0)(40,20)(0){0.9}
    \Text(60,30)[bc]{$(1-x)p_i$}  
    \Text(60,28)[tc]{$\alpha'$}  
    \Text(63,-25)[lb]{$q$}
    \Text(58,-35)[lt]{$a$}
    \Text(0,0)[]{$\widetilde{\bom{W}}(G'_i)$}
  \end{picture}
\end{array}
 =\dd  \gs\mu^\epsilon\,T^a_{\alpha'\alpha''}\;
\begin{array}{c}
  \begin{picture}(80,90)(-20,-45)
    \ArrowLine(17,20)(60,30)
    \GOval(0,0)(40,20)(0){0.9}
    \Text(60,32)[bc]{$p_i$}  
    \Text(60,27)[tc]{$\alpha''$}  
    \Text(0,0)[]{$\cal{M}$}
  \end{picture}
\end{array}
$
\caption{\label{fig:gauge-invariance-quark} Illustration of
Eq.~(\ref{Hquarkreduction}). The left hand side represents
$H_{\alpha'}^a(x;p_1,\dots,p_m)$ expressed in terms of 
$\widetilde{\bom{V}}$, with the gluon line contracted with $q$
indicated by the curly line with an arrowhead. The sum is originally
over graphs $G_i \in C'_+$, but this is the negative of the sum over
graphs $G_i \in C'_-$, in which the gluon connects to line $i$, as
depicted in the right hand side of the first equality. The remaining
Green function is $\widetilde{\bom{W}}$. The $q^\mu$ insertion then
cancels the adjoining propagator. After summing over graphs, we are left
with a color matrix times the complete $m$ particle tree amplitude.}
}

We now note that gauge invariance says something about $H$. We have
inserted a gluon line carrying momentum $q$ almost everywhere into tree
graphs with $m$ legs. The gluon has polarization $q^\mu$. Gauge
invariance tells us that if we inserted this gluon everywhere, that is if
we summed over the entire set of graphs $C'$, we would get zero. However
we have summed only over the graphs in $C'_+$, leaving out the graphs in
$C'_-$. Thus $H$ equals the negative of the sum over graphs in $C'_-$.
This enables us to calculate $H$ as follows:
\beqn
&&H_{\alpha'}^a(x;p_1,\dots,p_m)
\nonumber\\&&\qquad
=
-\lim_{q \to x p_i}
\sum_{G_i \in C'_-}
(1-x)^{-1/2}
q_\mu\,
\bar{u}((1-x) p_i)
\big[
\widetilde{\bom{V}}^\mu\!(G_i,
p_1,\dots,(1-x)p_i,\dots,p_m,q)
\big]_{\alpha'}^a
\nonumber\\&&\qquad=
\lim_{q \to x p_i}
\sum_{G_i' \in C_0}
(1-x)^{-1/2}\,
\gs\mu^{\epsilon}
T^a_{\alpha'\alpha''}\,
\frac{\bar{u}((1-x) p_i)\s{q}((1-x)\s{p}_i + \s{q})}
{((1-x)p_i + q)^2+\mi0}\
\nonumber\\&&
\hskip 4 cm \times
\big[
\widetilde{\bom{W}}(G_i'
,p_1,\dots,(1-x)p_i + q,\dots,p_m)
\big]_{\alpha''}
\nonumber\\
&&\qquad=
\sum_{G_i' \in C_0}
\gs\mu^{\epsilon}
T^a_{\alpha'\alpha''}\,
\bar{u}(p_i)\
\big[
\widetilde{\bom{W}}(G_i'
,p_1,\dots,p_i,\dots,p_m)
\big]_{\alpha''}
\nonumber\\
&&\qquad=
\gs\mu^{\epsilon}
T^a_{\alpha'\alpha''}\,
{\cal M}(p_1,\dots,p_i,\dots,p_m)
^{\alpha''}
\;\;.\ \ \
\label{Hquarkreduction}
\eeqn
Following the second equals sign, we have displayed the vertex at which
the gluon couples to the quark line $i$, together with the adjacent
propagator. Everything else we call $\widetilde{\bom{W}}$. Note that
$\widetilde{\bom{W}}$ is the Green function for an amputated
tree graph $G_i'$ with $m$ external partons. The partons have flavors
$\{f_1, \dots, f_m\}$, just as with our original loop graphs, and they
have the same momenta as the original partons except that parton $i$
carries momentum $(1-x)p_i + q$. Let us denote the set of all such
graphs $C_0$. The sum over $G_i \in C'_-$ implies that  $G_i'$ runs over
all of $C_0$. In the next step, we replace $\s{q}$ by $((1-x)\s{p}_i +
\s{q})$ next to the spinor. This gives a factor $((1-x)p_i + q)^2/
((1-x)p_i + q)^2 = 1$. After this cancellation, it is safe to take the
limit $q \to x p_i$. Also, we replace $\bar u((1-x)p_i)$ by $\sqrt{1-x}\,
\bar u(p_i)$. In the last step, we recognize that we have the complete
three level amplitude ${\cal M}$ for the $m$ external particles. This
calculation is illustrated in Fig.~\ref{fig:gauge-invariance-quark}.

We can now insert this result into Eq.~(\ref{collinearsubtractionfromH})
and perform the loop integral to obtain
\beqn
&&\sum_{G,i\in I_C(G)} {\cal C}_i(G;p_1,\dots,p_m)^\alpha
\nonumber\\
&&\qquad\quad=
\int\! \frac{ d^d l_j}{(2\pi)^d}\
\frac{f_{UV}(l_j, l_j - p_i)}{(l_j^2  + \mi0)((p_i - l_j)^2 + \mi0)}
\int_0^1\!\frac{dx}{x}
\ \delta\!\left(x - \frac{l_j\cdot n_i}{p_i\cdot n_i}\right)
2\mi\gs\mu^{\epsilon}\,
T^a_{\alpha\alpha'}
\nonumber\\
&&\qquad\qquad\qquad\qquad\times
\biggl\{
[(1-x) - 1]\
\gs\mu^{\epsilon}
T^a_{\alpha'\alpha''}\,
{\cal M}(p_1,\dots,p_i,\dots,p_m)^{\alpha''}
\biggr\}
\nonumber\\
&&\qquad\quad=
- 2\mi \gs^2 C_F\,\mu^{2\epsilon}
\int\! \frac{ d^d l_j}{(2\pi)^d}\
\frac{f_{UV}(l_j, l_j - p_i)}{(l_j^2  + \mi0)((p_i - l_j)^2 + \mi0)}
{\cal M}(p_1,\dots,p_i,\dots,p_m)^{\alpha}
\nonumber\\
&&\qquad\quad=
\frac{\as}{4\pi}C_F\frac{(4\pi)^\epsilon}{\Gamma(1-\epsilon)}
\left(-\frac 2\epsilon+{\cal O}(\epsilon)\right)\,
{\cal M}(p_1,\dots,p_i,\dots,p_m)^{\alpha}
\;\;.
\label{quarkcollinearsubtractions}
\eeqn
Thus, when we sum over all graphs the collinear subtraction terms
associated with an external quark line with label $i$, we get a simple
singular factor times the tree level amplitude ${\cal M}$.

For an external antiquark line we get the same result from essentially
the same calculation. We just have to exchange $u(p_i)$ for $\bar
v(p_i)$, reverse the order of matrix multiplications in Dirac spinor
space, and replace $\s{k} \to -\s{k}$ in the Dirac propagator
numerators.  

\subsection{\label{sec:collineargluon}Gluon line}

In this subsection we calculate the sum over graphs of the collinear
subtractions $\big| {\cal C}_i(G)\big\ra$ in the case that
the external line $i$ is a gluon line. There are two possibilities. Either
the external gluon connects to a virtual quark and antiquark, or else it
connects to two virtual gluons.

The case in which the external gluon connects to a virtual quark and
antiquark is simple. In this case the collinear limit of the numerator is
proportional to
\beqn
x(1-x) 
\s{p}_i\s{\varepsilon}(p_i)\s{p}_i
= x(1-x)\left[-p_i^2\s{\varepsilon}(p_i) 
+ 2\s{p}_i \,p_i\!\cdot\!{\varepsilon}(p_i)\right] = 0\;\;,
\eeqn
where ${\varepsilon}^\mu(p)$ is the polarization vector of the
external gluon. The first term is zero because of the $p_i^2 =0$.  The
second term is also zero because we consider only physical
polarizations of the external gluon, so that $p_i\cdot
{\varepsilon}(p_i) = 0$. Thus $\big| {\cal
  C}_i(G;p_1,\dots,p_m)\big\ra$ is zero when the external gluon
connects to a virtual quark and antiquark.

\FIGURE[ht]{
$\dd
\lim_{l_j\to xp_i} l_j^2 (l_j-p_i)^2 
\begin{array}{c}
  \begin{picture}(100,80)(-20,-40)
   \LongArrowArc(14,0)(17,-45,45)
    \Vertex(50,0){1}
    \Gluon(13.8,29)(50,0){4}{7}
    \Gluon(50,0)(80,0){4}{4.5}
    \Gluon(50,0)(13.8,-29){4}{7}
    \GOval(0,0)(40,20)(0){0.9}
    \Text(30,-25)[lt]{\small $l_j$}  
    \Text(32,20)[lb]{\small $l_{j}-p_i$}
   \Text(80,7)[b]{$p_i$}  
   \Text(80,-5)[t]{$b$}  
   \Text(0,0)[]{$\widetilde{\bom{V}}(G_i)$}
  \end{picture}
\end{array}$
\hfill~\\
~ 
\hfill
$~\qquad\quad \dd = \;\; \mi\gs\mu^{2\epsilon} 
\frac{\dd 2-x}{\dd x}\,T_{bc}^a\;
\begin{array}{c}
  \begin{picture}(100,90)(-20,-40)
    \Gluon(17,20)(60,30){4}{7}
    \ArrowGluonn(18,-20)(60,-30){4}{6}
    \GOval(0,0)(40,20)(0){0.9}
    \Text(60,35)[bc]{$(1-x)p_i$}  
    \Text(60,25)[tc]{$c$}  
    \Text(58,-25)[lb]{$xp_i$}
    \Text(58,-35)[lt]{$a$}
    \Text(0,0)[]{$\widetilde{\bom{V}}(G_i)$}
  \end{picture}
\end{array} 
 + \;\;\;
\dd \mi\gs\mu^{2\epsilon}
\frac{\dd 1+x}{\dd 1-x}\,T_{ba}^c\;
\begin{array}{c}
  \begin{picture}(100,90)(-20,-40)
    \ArrowGluonn(18,20)(60,30){4}{6}
    \Gluon(17,-20)(60,-30){4}{7}
    \GOval(0,0)(40,20)(0){0.9}
    \Text(60,35)[bc]{$(1-x)p_i$}  
    \Text(60,25)[tc]{$c$}  
    \Text(58,-25)[lb]{$xp_i$}
    \Text(58,-35)[lt]{$a$}
    \Text(0,0)[]{$\widetilde{\bom{V}}(G_i)$}
  \end{picture}
\end{array}
$
\caption{\label{fig:collgluon}
Illustration of Eq.~(\ref{collinearcoefficient1g}). The left hand side
represents $f^{C,0}_i(G;x;p_1,\dots,p_m)^\alpha$, which is defined in
Eq.~(\ref{collinearcoefficient1}) as the collinear limit of
$l_j^2\,(l_{j}-p_i)^2$ times the integrand $\widetilde{\cal
G}(G;l_1,\dots,l_{n};p_1,\dots,p_m)^\alpha$. Here the integrand 
$\widetilde{\cal G}$ consists of a polarization vector
$\varepsilon^\sigma(p_i)$, a vertex and two propagators times an
amputated Green function $\widetilde{\bom{V}}$, as in
Eq.~(\ref{collinearglue0}). As in Fig.~\ref{fig:collquark}, the
illustration represents the integrand: an integration over the loop
momentum is {\it not} implied.  On the right hand side, there are two
terms. In each, one gluon line of $\widetilde{\bom{V}}$ is contracted with
its momentum vector $(x p_i)$ or $(1-x)p_i$ respectively, indicated by
the arrowhead. The other gluon line is contracted with the original
polarization vector $\varepsilon^\sigma(p_i)$, which is written as
$\varepsilon_\nu((1-x)p_i)$ in the first term and $\varepsilon_\mu(xp_i)$
in the first term. The remaining factors in the right hand side of
Eq.~(\ref{collinearcoefficient1g}) are indicated.}
}

We now turn to the case in which the external gluon connects to two
virtual gluons, $g + g \to g$. Our analysis is similar to that of the
previous subsection for $q + g \to q$. There are, however, some
subtleties, so we present the argument in some detail. 

By assumption, line $i$, carrying momentum $p_i$ out of the graph,
connects to a virtual loop in $G$ at a gluon-gluon-gluon vertex. We
suppose that the gluon lines in the loop have labels $j$ and $j+1$. Thus
the gluon line with label $j$ carries momentum $l_j$ into the vertex and
the gluon line with label $j+1$ carries momentum $-l_{j+1} = p_i - l_j$
into the vertex. We write the integral for the graph as
\beqn \nn
{\cal G}(G,p_1,\dots,p_m)_b &=& \int\!\frac{d^dl}{(2\pi)^d}\
\widetilde{\cal G}(G;l_1,\dots,l_n,p_1,\dots,p_m)_b
\\
&=& - \mi\gs\mu^{\epsilon}
T^a_{bc}
\int\! \frac{d^dl_j}{(2\pi)^d}\
\frac{\varepsilon^\sigma(p_i) V^{(3)}_{\sigma\nu\mu}(p_i,l_j-p_i,-l_j)}
{(l_j^2  + \mi0)((p_i - l_j)^2 + \mi0)}
\nonumber \\ &&\qquad\qquad\qquad\qquad\times
\big[
\widetilde{\bom{V}}(G_i,p_1,\dots,p_i-l_j,\dots,p_m,l_j)
\big]^{\mu\nu}_{ac}
\;\;.\qquad
\label{collinearglue0}
\eeqn
We display a gluon color index $b$ for line $i$ in ${\cal G}$, leaving the
other indices on ${\cal G}$ unwritten. The amplitude
$\widetilde{\bom{V}}$ has a color index $a$ corresponding to the gluon
$j$ in the loop and a  gluon color index $c$ corresponding to line
$j+1$ in the loop. There is an explicit color matrix $T^a_{bc} = i
f_{bac}$ connecting the colors of $\widetilde{\bom{V}}$ to the color of
${\cal G}$. The amplitude $\widetilde{\bom{V}}$ also carries vector
indices $\mu,\nu$ corresponding to the polarization of the gluon lines
$j,j+1$. The polarization vector $\epsilon^\sigma(p_i)$ represents the
final state gluon on line $i$. We have displayed the integration over the
loop momentum $l_j$, the propagators for the gluons in the loop and the
gluon-gluon-gluon vertex $V$,
\beqn
V_{\sigma\nu\mu}(p_i,l_j-p_i,-l_j)
=
g_{\sigma\nu} (l_j - 2p_i)_\mu 
+ g_{\nu\mu} (p_i - 2l_j)_\sigma
+ g_{\mu\sigma} (p_i + l_j)_\nu\;\;.
\eeqn
Everything else is included in the Feynman amplitude
$\widetilde{\bom{V}}$ for the amputated tree level 
graph $G_i$ obtained by omitting the propagators $j$ and $j+1$ and the
vertex that attaches these propagators to external line $i$ in graph
$G$. Thus the amplitude $\widetilde{\bom{V}}$ is defined by
Eq.~(\ref{collinearglue0}). The momentum arguments of
$\widetilde{\bom{V}}$ are the external momenta of the original graph,
plus the momenta $p_i-l_j$ and $l_j$ of the two gluon lines.

With the notation thus defined, we are ready to calculate. The
subtraction term $f^{C,0}$ defined in Eq.~(\ref{collinearcoefficient1})
can be expressed in terms of $\widetilde{\bom{V}}$ as
\beqn
\lefteqn{f^{C,0}_i(G;x;p_1,\dots,p_m)_b
=
\lim_{l_j \to x\, p_i}
l_j^2\,(p_i - l_j)^2\,
\widetilde{\cal G}(G;l_1,\dots,l_{n};p_1,\dots,p_m)_b}
\nonumber\\
&&\quad=
\lim_{l_j \to x\, p_i}
(-\mi\gs\mu^{\epsilon})
T^a_{bc}
\varepsilon^\sigma(p_i)\,
V_{\sigma\nu\mu}(p_i,l_j-p_i,-l_j)
\big[
\widetilde{\bom{V}}(G_i,p_1,\dots,p_i-l_j,\dots,p_m,l_j)
\big]^{\mu\nu}_{ac}
\nonumber\\
&&\quad =
 \mi\gs\mu^{\epsilon} T^a_{bc} \frac{2-x}{x}\ (x p_i)_\mu\,
\varepsilon_\nu((1-x)p_i)\,
\big[
\widetilde{\bom{V}}(G_i,p_1,\dots,(1-x)p_i,\dots,p_m,xp_i)
\big]^{\mu\nu}_{ac}
\nonumber\\
&& \qquad + \mi\gs\mu^{\epsilon}
T^c_{ba}
\frac{1+x}{1-x}\
((1-x) p_i)_\nu\,
\varepsilon_\mu(x p_i)\,
\big[
\widetilde{\bom{V}}(G_i,p_1,\dots,(1-x)p_i,\dots,p_m,xp_i)
\big]^{\mu\nu}_{ac}
\;\;.\qquad\;
\label{collinearcoefficient1g}
\eeqn
There are two terms here. In the first, we contract the polarization
index $\nu$ of the gluon line with momentum $(1-x)p_i$ into the index
$\nu$ of the polarization vector $\varepsilon_\nu(p_i) =
\varepsilon_\nu((1-x)p_i)$, while we contract the polarization index
$\mu$ of the gluon line with momentum $xp_i$ with the index $\mu$ of its
momentum vector $(xp_i)_\mu$. In the second term the situation is just
reversed, so that gluon line with index $\nu$ and momentum $(1-x)p_i$ is
contracted with its momentum vector. It is useful to represent these two
terms diagrammatically as different graphs, depicting the gluon line
contracted with the polarization vector as a curly line and the gluon
line contracted with its momentum vector as curly line with an arrowhead,
as illustrated in Fig.~\ref{fig:collgluon}. We can denote these two
graphs as $G_i^1$ and $G_i^2$.

In the case of the second term, we revise the notation, exchanging $\mu
\leftrightarrow \nu$, $ a \leftrightarrow  c$, $x \leftrightarrow (1-x)$
and exchanging the positions of the $xp_i$ and $(1-x)p_i$ arguments of
$\widetilde{\bom{V}}$. Then the two terms have the same form except that
in one term we have the amplitude $\widetilde{\bom{V}}(G^1_i,\dots)$ and
in the other we have $\widetilde{\bom{V}}(G^2_i,\dots)$

As in the previous subsection, we would now like to sum over graphs. We
consider one-loop amputated graphs $G$ with $m$ external particles having
flavors $\{f_1,\dots,f_m\}$ and momenta $\{p_1,\dots,p_m\}$. We are
considering the collinear subtraction for parton $i$, which we assume
here is a gluon. We consider only graphs that have a collinear
divergence for external leg $i$: $i\in I_C(G)$. We call this class of
graphs $C$.

The corresponding graphs $G_i^1$ and $G_i^2$ are amputated tree graphs
with $m+1$ external legs. The flavors of the first $m$ external particles
are $\{f_1,\dots,f_m\}$, the same as for the graph $G$. The momenta of
these particles are the same as for $G$ except for particle $i$, which
carries momentum $(1-x)p_i$ instead of $p_i$. The external particle with
index $m+1$ is a gluon with momentum $xp_i$. Let us call the set of all
such graphs $C'$. When we sum over all graphs $G \in C$, the graphs
$G_i^1$ together with the graphs $G_i^2$ include all graphs in $C'$ except
for the graphs in which gluon $m+1$ couples directly to gluon $i$. These
graphs are absent because graphs with a self-energy insertion on external
line $i$ are {\it not} included in $C$. Let us call the set of
($m+1$)-particle graphs that we do get $C'_+$ and the graphs that we
don't get
$C'_-$. 

The collinear subtraction defined in Eq.~(\ref{collinearsubtraction}),
summed over graphs $G \in C$, is
\beqn
\lefteqn{\sum_{G \in C} {\cal C}_i(G;p_1,\dots,p_m)_b}
\nonumber\\&=&
\int\! \frac{ d^d l_j}{(2\pi)^d}\
\frac{f_{UV}(l_j, l_j - p_i)}{(l_j^2  + \mi0)((p_i - l_j)^2 + \mi0)}
\int_0^1\!dx\ \delta\!\left(x - \frac{l_j\cdot n_i}{p_i\cdot n_i}\right) 
\sum_{G \in C}
f^C_i(G;x;p_1,\dots,p_m)_b
\nonumber\\
&=&
\int\! \frac{ d^d l_j}{(2\pi)^d}\
\frac{f_{UV}(l_j, l_j - p_i)}{(l_j^2  + \mi0)((p_i - l_j)^2 + \mi0)}
\int_0^1\!dx\
\delta\!\left(x - \frac{l_j\cdot n_i}{p_i\cdot n_i}\right)
\mi\gs\mu^{\epsilon}
T^a_{bc}
\nonumber\\
&&\times
\biggl\{
\frac{(2-x)}{x(1-x)}\,H_{ac}(x;p_1,\dots,p_m)
-
\frac{2}{x}\,H_{ac}(0,p_1,\dots,p_m)
-
\frac{H_{ac}(1,p_1,\dots,p_m)}{(1-x)}
\biggr\}
\;\;,
\label{collinearsubtractionfromHg}
\eeqn
where we define
\beqn
&&H_{ac}(x;p_1,\dots,p_m)
=
\nonumber\\
&&\hskip 0.5 cm
(1-x)
\lim_{q \to x p_i}
\sum_{G_i \in C'_+}
q_\mu\,
\varepsilon_\nu((1-x)p_i)\,
\big[
\widetilde{\bom{V}}(G_i,p_1,\dots,(1-x)p_i,\dots,p_m,q)
\big]^{\mu\nu}_{ac}
\;\;.\ \ \
\eeqn
\FIGURE[ht]{
$\dd-(1-x)
\sum_{G_i\in C'_-} \lim_{q\to xp_i}\;
\begin{array}{c}
  \begin{picture}(100,90)(-20,-45)
    \Gluon(17,20)(60,30){4}{7}
    \ArrowGluonn(18,-20)(60,-30){4}{6}
    \GOval(0,0)(40,20)(0){0.9}
    \Text(60,35)[bc]{$(1-x)p_i$}  
    \Text(60,25)[tc]{$c$}  
    \Text(58,-25)[lb]{$q$}
    \Text(58,-35)[lt]{$a$}
    \Text(0,0)[]{$\widetilde{\bom{V}}(G_i)$}
  \end{picture}
\end{array} 
$
\hfill~\\
~ 
\hfill 
$~\qquad\qquad\;\; =\dd -(1-x)
\lim_{q\to xp_i} 
\sum_{G'_i\in C_0}\;
\begin{array}{c}
  \begin{picture}(100,90)(-20,-45)
    \Gluon(60,30)(17,20){4}{7}
    \ArrowGluonn(37,20)(60,-30){4}{8}
    \GOval(0,0)(40,20)(0){0.9}
    \Text(60,35)[bc]{$(1-x)p_i$}  
    \Text(60,23)[tc]{$c$}  
    \Text(63,-25)[lb]{$q$}
    \Text(58,-35)[lt]{$a$}
    \Text(0,0)[]{$\widetilde{\bom{W}}(G'_i)$}
  \end{picture}
\end{array}
 =\dd  (1-x) \gs\mu^\epsilon\,T^a_{cd}\;
\begin{array}{c}
  \begin{picture}(80,90)(-20,-45)
    \Gluon(17,20)(60,30){4}{7}
    \GOval(0,0)(40,20)(0){0.9}
    \Text(60,35)[bc]{$p_i$}  
    \Text(60,25)[tc]{$d$}  
    \Text(0,0)[]{$\cal{M}$}
  \end{picture}
\end{array}
$
\caption{\label{fig:gauge-invariance-gluon}
Illustration of Eq.~(\ref{Hgluonreduction}). The left hand side represents
$H_{ac}(x;p_1,\dots,p_m)$ expressed in terms of 
$\widetilde{\bom{V}}$, with the gluon line contracted with $q_\mi$
indicated by the curly line with an arrowhead. The sum is originally
over graphs $G_i \in C'_+$, but this is the negative of the sum over
graphs $G_i \in C'_-$, in which the gluon connects to line $i$, as
depicted in the right hand side of the first equality. The remaining
Green function is $\widetilde{\bom{W}}$. The $q_\mu$ insertion then gives
two terms, not depicted in the figure. One of the terms cancels when
summed over graphs, while in the other the propagator adjoining the
$q_\mu$ insertion is cancelled. After summing over graphs, we are left
with a color matrix times the complete $m$ particle tree amplitude.}
}

We now use gauge invariance. We have inserted a gluon line carrying
momentum $q$ with polarization $q^\mu$ almost everywhere into tree graphs
with $m$ legs. Gauge invariance tells us that if we inserted this gluon
everywhere, that is if we summed over the entire set of graphs $C'$, we
would get zero. However we have summed only over the graphs in $C'_+$,
leaving out the graphs in $C'_-$. Thus $H$ equals the negative of the sum
over graphs in $C'_-$. This enables us to calculate $H$ as follows:
\beqn
\lefteqn{
H_{ac}(x;p_1,\dots,p_m)
}
\nonumber\\
&=&
-(1-x)
\lim_{q \to x p_i}
\sum_{G_i\in C'_-}
q_\mu\,
\varepsilon_\nu((1-x)p_i)\,
\big[
\widetilde{\bom{V}}(G_i,p_1,\dots,(1-x)p_i,\dots,p_m,q)
\big]^{\mu\nu}_{ac}
\nonumber\\&=&
-(1-x)
\lim_{q \to x p_i}
\sum_{G'_i \in C_0}
\gs\mu^{\epsilon}
T^a_{cd}\,
\frac{q_\mu
\varepsilon_\nu((1-x)p_i)V^{\nu\lambda\mu}((1-x)p_i,-(1-x)p_i-q,q)}
{((1-x)p_i + q)^2+\mi0}
\nonumber\\
&&\times
\big[
\widetilde{\bom{W}}(G'_i
,p_1,\dots,(1-x)p_i + q,\dots,p_m)
\big]_{\lambda,d}
\nonumber\\
&=&
(1-x)
\lim_{q \to x p_i}
\sum_{G'_i\in C_0}
\Biggl\{
\gs\mu^{\epsilon}
T^a_{cd}\,
\varepsilon_\nu((1-x)p_i)\
\big[
\widetilde{\bom{W}}(G'_i
,p_1,\dots,(1-x)p_i + q,\dots,p_m)
\big]^\nu_{d}
\nonumber\\&& -
\gs\mu^{\epsilon}
T^a_{cd}\,
\frac{q\cdot\varepsilon((1-x)p_i)\,((1-x)p_i + q)_\lambda}
{((1-x)p_i + q)^2+\mi0}
\widetilde{\bom{W}}(G'_i
,p_1,\dots,(1-x)p_i + q,\dots,p_m)
\big]^\lambda_{d}
\biggr\}
\nonumber\\
&=&
(1-x)\
\gs\mu^{\epsilon}
T^a_{cd}\,
\varepsilon_\nu(p_i)\
\sum_{G'_i\in C_0}
\big[
\widetilde{\bom{W}}(G'_i
,p_1,\dots,p_i,\dots,p_m)
\big]^\nu_{d}
\nonumber\\
&=&
(1-x)\
\gs\mu^{\epsilon}
T^a_{cd}\,
{\cal M}(p_1,\dots,p_i,\dots,p_m)^{d}
\;\;.\ \ \
\label{Hgluonreduction}
\eeqn
After the second equals sign, we have displayed the vertex at which the
gluon with momentum $q$ couples to the gluon line $i$, together with the
adjacent propagator. This multiplies the Green function
$\widetilde{\bom{W}}$ for an amputated tree graph $G_i'$ with $m$
external partons. The partons have the same flavors as in our original
loop graphs, and they have the same momenta except that parton $i$
carries momentum $(1-x)p_i + q$. We denote the set of all such graphs
$C_0$. The sum over $G_i \in C'_-$ implies that  $G_i'$ runs over all of
$C_0$. The vector and color indices displayed for $\widetilde{\bom{W}}$
are those of the line $i$. Now, when we evaluate $q_\mu$ contracted with
the vertex function and use $(1-x)p_i \cdot \varepsilon((1-x)p_i) = 0$ as
well as
$p_i^2 = 0$, we get two terms, as indicated on the right hand side of the
third equality. In the second term, the amplitude  $\widetilde{\bom{W}}$
is contracted with the momentum carried by the gluon line. After summing
over graphs  $G'_i$, this term vanishes.   Finally, we take the limit $q
\to x p_i$ and recognize that we have the complete three level amplitude
${\cal M}$ for $m$ external particles. This calculation is illustrated in
Fig.~\ref{fig:gauge-invariance-gluon}.

We can now insert this result into Eq.~(\ref{collinearsubtractionfromHg})
and perform the loop integral to obtain
\beqn
\lefteqn{\sum_{G \in C} {\cal C}_i(G;p_1,\dots,p_m)^b}
\nonumber\\
&=&
\int\! \frac{ d^d l_j}{(2\pi)^d}\
\frac{f_{UV}(l_j, l_j - p_i)}{(l_j^2  + \mi0)((p_i - l_j)^2 + \mi0)}
\int_0^1\!dx
\ \delta\!\left(x - \frac{l_j\cdot n_i}{p_i\cdot n_i}\right)
\mi\gs\,\mu^{\epsilon}
T^a_{bc}
\nonumber\\
&&\times
\biggl\{
\left[ 
\frac{(2-x)}{x}
-\frac{2}{x}
-\frac{0}{1-x}\right]\
\gs\mu^{\epsilon}
T^a_{cd}\,
{\cal M}(p_1,\dots,p_i,\dots,p_m)^{d}
\biggr\}
\nonumber\\
&=&
-\mi \gs^2 C_A\,\mu^{2\epsilon}
\int\! \frac{ d^d l_j}{(2\pi)^d}\
\frac{f_{UV}(l_j, l_j - p_i)}{(l_j^2  + \mi0)((p_i - l_j)^2 + \mi0)}
{\cal M}(p_1,\dots,p_i,\dots,p_m)^{b}
\nonumber\\
&=&
\frac{\as}{4\pi}C_A\frac{(4\pi)^\epsilon}{\Gamma(1-\epsilon)}
\left(-\frac{1}\epsilon+{\cal O}(\epsilon)\right)\,
{\cal M}(p_1,\dots,p_i,\dots,p_m)^{b}
\;\;.
\label{gluoncollinearsubtractions}
\eeqn
Thus, when we sum over all graphs the collinear subtraction terms
associated with an external gluon line with label $i$, we get a simple
singular factor times the tree level amplitude ${\cal M}$.

\section{\label{sec:final}Final formulas}

There are two steps to the algorithm that we have outlined here for
generating the expressions to be used in a numerical calculation of the
one-loop graphs for a QCD amplitude.

The first step applies graph by graph, generating subtractions for each
graph. If the graph is ultraviolet divergent, one generates a subtraction
term for the integrand of the graph. The subtraction term matches the
integrand when the loop momentum is much larger than the external
momenta. In this way, the ultraviolet divergence of the integral is
removed. The subtraction terms are designed to have the same effect as
standard $\overline{\rm MS}$ renormalization. If the graph has soft or
collinear divergences, or both, then one generates corresponding
subtraction terms. The subtraction terms match the integrand in the soft
and collinear limits of the loop momentum. In this way, the soft and
collinear divergences of the integral are removed.

The second step generates a next-to-leading order contribution to the
$m$-parton amplitude that is proportional to the tree level matrix element
(summed over tree graphs). This contribution has the form
\beqn
\bom{\mathrm{I}}^V\!(\epsilon)
\big|{\cal M}_{\rm tree}\left(p_1,\dots,p_m\right)\big\ra\;\;,
\eeqn  
where $\bom{\mathrm{I}}^V(\epsilon)$ is a singular function of the
dimensional regularization parameter $\epsilon$ and is a linear operator
on the color space of the amplitude $\big|{\cal M}_{\rm
tree}\left(p_1,\dots,p_m\right)\big\ra$. There are three parts to 
$\bom{\mathrm{I}}^V(\epsilon)$. The first comes from the factors
$\sqrt{r_i}$ for each external line that relate the scattering matrix to the
amputated Green function. These contributions are given in
Eq.~(\ref{externallegs}). There is one term for each external line
$i$. The second contribution comes from adding back the collinear
subtractions. Again there is one contribution for each external line, as
given in Eq.~(\ref{quarkcollinearsubtractions}) for quarks and
Eq.~(\ref{gluoncollinearsubtractions}) for gluons. Finally, there is a
third contribution that comes from adding back the soft subtractions.
There is one term for each pair of external lines, as given in
Eq.~(\ref{softcorrections}). The net result is
\beqn
\bom{\mathrm{I}}^V\!(\epsilon) = 
\frac{\alpha_s}{4 \pi}\,
\frac{(4\pi)^\epsilon}{\Gamma(1-\epsilon)}
\Bigg(\frac{1}{\epsilon^2}
\sum_{
  \begin{subarray}{c}
    i,j = 1\\
    i\neq j
    \end{subarray}
}^m \bom{T}_i\cdot\bom{T}_j 
\left(\frac{\mu^2}{-2p_i\!\cdot\!p_j}\right)^{\epsilon}
-\frac{1}{\epsilon} \sum_{i=1}^m \gamma_i
\Bigg)\;\;,
\label{singularV}
\eeqn
where the $\gamma_i$ factors are
\beqn
\gamma_q=\gamma_{\bar{q}} = \frac32C_F\;\;, \qquad 
\gamma_g = \frac{11}6C_A-\frac46T_Rn_f\;\;.
\eeqn
The $1/\epsilon^2$ and $1/\epsilon$ poles are all of infrared origin.

We now turn to the complete calculation of an infrared safe cross section
$\sigma$, which we write in the form (following as much as possible the
notation of \cite{Catani:1997vz})
\beqn
\sigma =
\sigma^{{\rm LO}\{m -n_I\}}
+\sigma^{{\rm NLO}\{m+1-n_I\}}
+ \sigma^{{\rm NLO}\{m-n_I\}}\;\;.
\eeqn  
There are a lowest order term and two next-to-leading order terms.
We suppose that there are $n_I$ initial state particles. (In applications,
$n_I$ is 0, 1, or 2.) In the  LO term there are $m$ total particles,
$n_I$ in the initial state and the remaining $m - n_I$ in the final
state. We denote this contribution by $\sigma^{{\rm LO}\{m -n_I\}}$, but
do not discuss its calculation. The NLO term from real parton emission
has $m+1-n_I$ final state particles. This is calculated from the
$m+1-n_I$ particle matrix element minus subtractions. One can use, for
example, the subtraction scheme proposed by Catani and Seymour
\cite{Catani:1997vz}. We denote this contribution by $\sigma^{{\rm
NLO}\{m+1-n_I\}}$, but do not further discuss its calculation. The
remaining NLO term has $m - n_I$ final state particles. It is calculated
as follows:
\beqn
\sigma^{{\rm NLO}\{m-n_I\}}&=&
\sum_G \int d\Phi^{m-n_I}\
F_J^{(m-n_I)}(p_1,\dots,p_m)\,
\int\! \frac{d^4 l}{(2 \pi)^4}
\nonumber\\
&&\qquad\times 2\,{\rm Re}\biggl\{
\big\la{\cal M}_{\rm tree}\left(p_1,\dots,p_m\right)\big|
\widetilde{\cal G}(G;\{l\};p_1,\dots,p_m)\big\ra
\nonumber\\
&&\qquad\qquad\qquad-
\big\la{\cal M}_{\rm tree}\left(p_1,\dots,p_m\right)\big|
\widetilde{\cal R}(G;\{l\};p_1,\dots,p_m)\big\ra
\nonumber\\
&&\qquad\qquad\qquad-
\sum_{\{i,j\} \in I_S(G)}
\big\la{\cal M}_{\rm tree}\left(p_1,\dots,p_m\right)\big|
\widetilde{\cal S}_{ij}(G;\{l\};p_1,\dots,p_m)\big\ra
\nonumber\\
&&\qquad\qquad\qquad-
\sum_{i\in I_C(G)}
\big\la{\cal M}_{\rm tree}\left(p_1,\dots,p_m\right)\big|
\widetilde{\cal C}_{i}(G;\{l\};p_1,\dots,p_m)\big\ra
\biggr\}
\nonumber\\
&+&\int d\Phi^{m-n_I}\
F_J^{(m-n_I)}(p_1,\dots,p_m)\,
\nonumber\\
&&\quad\times
2\lim_{\epsilon \to 0}
\big\la{\cal M}_{\rm tree}\left(p_1,\dots,p_m\right)\big|
\bom{\mathrm{I}}^V\!(\epsilon)
+ \bom{\mathrm{I}}^R(\epsilon)
\big|{\cal M}_{\rm tree}\left(p_1,\dots,p_m\right)\big\ra
\;\;.
\eeqn  
In the first term here, there is a sum over amputated one-loop graphs
$G$. We integrate over the phase space $d\Phi$ for $m - n_I$ final state
particles with momenta $p_{n_I+1},\dots p_m$, where the initial state
particles have momenta $p_1,\dots,p_{n_I}$. Next, we supply a measurement
$F_J$ appropriate to the infrared safe observable that we wish to
calculate. (``J'' is for ``jet''.) Then there is an integration over the
loop momentum $l$ as described in Sec.~\ref{subsec:oneloop}. 
Now there follow four
terms inside the loop integration. The first is the integrand
$\widetilde{\cal G}$ for the graph $G$ times the complex conjugate of the
tree amplitude for $m-n_I$ final state particles. The second is
$\big\la{\cal M}_{\rm tree}\big|$ times the renormalization subtraction
${\cal R}$, which is zero if the virtual loop graph is ultraviolet
convergent. The renormalization subtraction is defined in
Sec.~\ref{sec:UV}.  The third term contains the soft gluon
subtractions ${\cal S}$ for graph $G$, one subtraction term for each pair
$\{i,j\}$ of external lines that corresponds to a soft divergence. The
soft subtraction is defined in Sec.~\ref{sec:soft}. The fourth term
contains the collinear subtractions ${\cal C}$ for graph $G$,  one
subtraction term for each external line $i$ that corresponds to a
collinear divergence. The collinear subtraction is defined in
Sec.~\ref{sec:collinear}.

There is one final term, which has no integral over a loop momentum and
involves only the tree amplitude. Here we have the singular function
$\bom{\mathrm{I}}^V\!(\epsilon)$  from Eq.~(\ref{singularV}) that adds
back the soft and collinear subtractions from the virtual graphs. We
also have a singular function $\bom{\mathrm{I}}^R(\epsilon)$ that adds
back the soft and collinear subtractions from the real graphs. Naturally,
exactly what function goes here depends on what subtraction scheme one
uses for the real graphs. However, the $1/\epsilon$ and $1/\epsilon^2$
poles, whose structure follows from the structure of QCD, cancel between
$\bom{\mathrm{I}}^V\!(\epsilon)$ and $\bom{\mathrm{I}}^R(\epsilon)$.
(For instance, one can easily check that this cancellation works for the
scheme of \cite{Catani:1997vz}, where $\bom{I}(\epsilon)$ corresponds to 
our $2\,\bom{I}^R(\epsilon)$.) Generally a finite contribution remains
in the limit $\epsilon \to 0$.

\acknowledgments{
We thank Stefano Catani for help with the soft gluon subtraction. This
work was supported in part by US Department of Energy, contract
DE-FG0396ER40969 as well as by the Hungarian Scientific Research
Fund grant OTKA T-038240.
}

\appendix
\section{\label{app:renorm} UV counterterms}

In this appendix we give the ultraviolet counterterms for the one-loop
propagator and vertex corrections in the massless case. We follow the
general prescription given in Sec.~\ref{sec:UV}, which works in every
case except for the gluon self-energy, where we need an extra subtraction
that was not discussed in Sec.~\ref{sec:UV} in order to handle the
quadratic divergence.

Since the divergent graphs and the corresponding counterterms
depend on the form of the Feynman rules it is useful to define them. In
Feynman gauge we have the following rules:
\begin{itemize}
\item Gluon propagator:
  \beqn
  \mi D_{ab}^{\mu\nu}(k) =
  -\frac{\mi g_{\mu\nu}}{k^2 + \mi 0}\, \delta_{ab}\;\;,
  \eeqn
  where $a$ and $b$ are the color indices and $k$ is the momentum
  carried by this line. 
\item Quark propagator:
  \beqn
  \mi S_{\alpha\beta}(p) = 
  \frac{\mi\s{p}}{p^2+ \mi 0}\, \delta_{\alpha\beta}\;\;,
  \eeqn
  where $\alpha$, $\beta$ are the color indices and $p$ is the momentum
  carried by this line. 
\item Ghost propagator:
  \beqn
  \mi\widetilde{D}_{ab}(k) =
  \frac{{\mathrm{i}}\delta_{ab}}{k^2+ \mi 0}\;\;,
  \eeqn
  where $a$, $b$ are the color indices and $k$ is the momentum
  carried by this line. 
\item Gluon-quark vertex: 
  \beqn
  \Gamma^\mu_{\beta\alpha} = 
  {\mathrm{i}}\gs\mu^{\epsilon}\gamma^\mu t^a_{\beta\alpha}\;\;,
  \eeqn
  where $t^a_{\beta\alpha}$ is the fundamental representation of the
  generators of the gauge group and $a$, $\alpha$ and $\beta$ are the color
  indices of the gluon, antiquark and quark respectively.
\item Three-gluon vertex:
  \beqn
  G_{a_1a_2a_3}^{\mu_1\mu_2\mu_3}(p_1,p_2,p_3) = 
  \mi\gs\mu^{\epsilon} F^{a_1}_{a_2a_3}
  V^{\mu_1\mu_2\mu_3}(p_1,p_2,p_3)\;\;,
  \eeqn
  where $F^{a_1}_{a_2a_3} = -\mi f_{a_1a_2a_3}$ is the adjoint 
  representation of the generators of the gauge group and the kinetic part of
  the three-gluon vertex is
  \beqn
  V^{\mu_1\mu_2\mu_3}(p_1,p_2,p_3) = g^{\mu_3\mu_1}(p_1-p_3)^{\mu_2}
  + g^{\mu_2\mu_3}(p_3-p_2)^{\mu_1} +
  g^{\mu_1\mu_2}(p_2-p_1)^{\mu_3}\;\;, \qquad
  \eeqn
  where the $p_i$ momenta are outgoing.  
\item Four-gluon vertex:
  \beqn\nn
  W^{\mu_1\mu_2\mu_3\mu_4}_{a_1a_2a_3a_4} &=& \mi\gs^2\mu^{2\epsilon} 
  \big(
  F^b_{a_1a_2}F^b_{a_3a_4}
  (g^{\mu_1\mu_3} g^{\mu_2\mu_4}-g^{\mu_1\mu_4}g^{\mu_2\mu_3})\\\nn
  &&\qquad\quad+F^b_{a_1a_4}F^b_{a_2a_3}
  (g^{\mu_1\mu_2}g^{\mu_3\mu_4}-g^{\mu_1\mu_3}g^{\mu_2\mu_4})\\
  &&\qquad\quad+F^b_{a_1a_3}F^b_{a_2a_4}
  (g^{\mu_1\mu_2}g^{\mu_3\mu_4}-g^{\mu_1\mu_4}g^{\mu_2\mu_3})
  \big)\;\;,
  \eeqn
  where the $a_i$ are color indices.
\item Gluon-ghost vertex:
  \beqn
  \widetilde{G}^\mu_{abc}(k) = \mi\gs\mu^{\epsilon} F^a_{bc} k^\mu\;\;,
  \eeqn
  where $a$, $b$, and $c$ are the color indices of the gluon, ghost, and
  antighost legs respectively and $k$ is the momentum of the outgoing 
  ghost.
\end{itemize}

We use the standard notation for color factors, $C_A = N_c$ ( = 3 for
SU(3)), $C_F = (N_c^2 -1)/(2N_c)$, $T_R = 1/2$.

\subsection*{Quark self-energy}

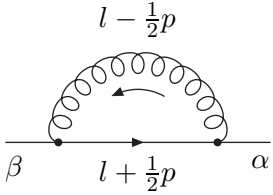
\FIGURE[hl]{
\begin{picture}(110,75)(-5,-15)
  \Vertex(80,10){1.5} 
  \Vertex(20,10){1.5}
  \ArrowLine(0,10)(100,10)
  \LongArrowArc(50,10)(20,60,120)
  \GlueArc(50,10)(30,0,180){4}{11}   
  \Text(50,5)[ct]{$l+\frac{1}{2} p$}
  \Text(50,50)[cb]{$l-\frac{1}{2} p$}
  \Text(100,5)[rt]{$\alpha$}
  \Text(0,5)[lt]{$\beta$}
\end{picture}
\caption{\label{fig:quarkselfenergy}Quark self-energy graph.}
}
The quark self-energy graph is the simplest one-loop Green function.  
It has logarithmic and linear UV divergences and we can easily define
the UV  counterterm by applying the prescription introduced in
Sec.~\ref{sec:UV}.

This graph is shown in Fig.~\ref{fig:quarkselfenergy} and the
renormalized quark self-energy is defined by the following integral: 
\beqn\label{renormquarkprop}
 \Sigma^R_{\alpha\beta}(p) &=& 
\int\frac{d^dl}{(2\pi)^d}
\left[\widetilde{\Sigma}_{\alpha\beta}(l, p)
  - \widetilde{\Sigma}_{\alpha\beta}^{UV}(l,p)\right]
\;\;,
\eeqn
where the integrand of the self-energy graph is given by
\beqn
\widetilde{\Sigma}_{\alpha\beta}(l,p) &=&  
-\mi\gs^2 \mu^{2\epsilon} C_F \delta_{\alpha\beta}\,
\frac{\gamma^\nu\left(\s l+\frac{1}{2} \s p\right)\gamma_\nu}
{\left(\left(l+\frac{1}{2} p\right)^2+\mi0\right)
 \left(\left(l-\frac{1}{2} p\right)^2+\mi0\right)}\;\;,
\eeqn
and the corresponding ultra-violet counterterm is
\beqn\label{quarkselfUVcounter}
\widetilde{\Sigma}_{\alpha\beta}^{UV}(l,p) = 
 -\mi\gs^2 \mu^{2\epsilon} C_F \delta_{\alpha\beta} 
\frac{\gamma^\nu\left(\s l + \frac{1}{2} \s p\right)\gamma_\nu}
{\left(l^2-\mu^2 e^{-1}+\mi0\right)^2}\;\;.
\eeqn
Here $e = 2.71828\dots\,$ is the base of natural logarithms.

\subsection*{Gluon self-energy}

\FIGURE[ht]{
$\begin{array}{c}
\begin{picture}(100,80)(0,-40)
  \Vertex(75,0){1.5} 
  \Vertex(25,0){1.5}
  \GlueArc(50,0)(25,0,180){4}{8}
  \GlueArc(50,0)(25,180,360){4}{8}
  \Gluon(0,0)(25,0){3.5}{3}
  \Gluon(75,0)(100,0){3.5}{3}
  \Text(50,30)[bc]{$l_2$}
  \Text(50,-33)[tc]{$l_1$}
  \Text(0,-7)[tl]{$\mu$, $a$}
  \Text(100,-7)[tr]{$\nu$, $b$}
\end{picture}
\end{array}
\;\; \phantom{+} \;\;
\begin{array}{c}
\begin{picture}(100,80)(0,-40)
  \Vertex(75,0){1.5} 
  \Vertex(25,0){1.5}
  \ArrowArcn(50,0)(25,0,180)
  \ArrowArcn(50,0)(25,180,360)
  \LongArrowArcn(50,0)(20,120,60)
  \LongArrowArcn(50,0)(20,-60,-120)
  \Gluon(0,0)(25,0){4}{3}
  \Gluon(75,0)(100,0){4}{3}
  \Text(50,30)[bc]{$l_2$}
  \Text(50,-30)[tc]{$l_1$}
  \Text(0,-7)[tl]{$\mu$, $a$}
  \Text(100,-7)[tr]{$\nu$, $b$}
\end{picture}
\end{array}
$
$\dd\frac{\dd 1}{\dd 2} \left[
\begin{array}{c}
\begin{picture}(100,80)(0,-40)
  \Vertex(75,0){1.5} 
  \Vertex(25,0){1.5}
  \DashArrowArcn(50,0)(25,0,180){2}
  \DashArrowArcn(50,0)(25,180,360){2}
  \LongArrowArcn(50,0)(20,120,60)
  \LongArrowArcn(50,0)(20,-60,-120)
  \Gluon(0,0)(25,0){4}{3}
  \Gluon(75,0)(100,0){4}{3}
  \Text(50,30)[bc]{$l_2$}
  \Text(50,-30)[tc]{$l_1$}
  \Text(0,-7)[tl]{$\mu$, $a$}
  \Text(100,-7)[tr]{$\nu$, $b$}
\end{picture}
\end{array}
\;\; + \;\;
\begin{array}{c}
\begin{picture}(100,80)(0,-40)
  \Vertex(75,0){1.5} 
  \Vertex(25,0){1.5}
  \DashArrowArc(50,0)(25,0,180){2}
  \DashArrowArc(50,0)(25,180,360){2}
  \LongArrowArcn(50,0)(20,120,60)
  \LongArrowArcn(50,0)(20,-60,-120)
  \Gluon(0,0)(25,0){4}{3}
  \Gluon(75,0)(100,0){4}{3}
  \Text(50,30)[bc]{$l_2$}
  \Text(50,-30)[tc]{$l_1$}
  \Text(0,-7)[tl]{$\mu$, $a$}
  \Text(100,-7)[tr]{$\nu$, $b$}
\end{picture}
\end{array}
\right]$
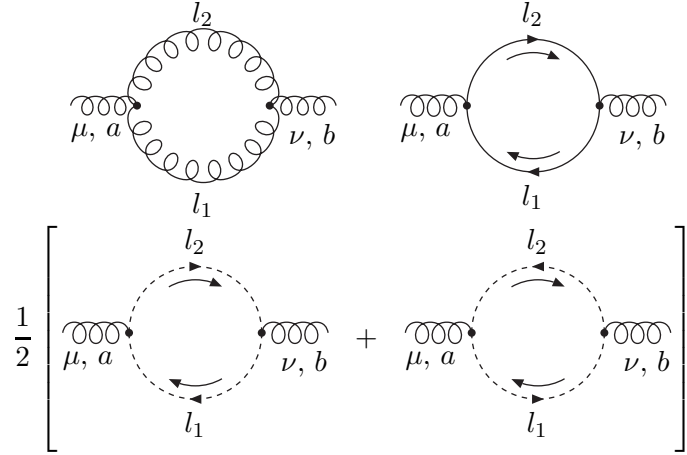
\caption{\label{fig:gluonself}Feynman graphs that contribute to the
  gluon self-energy.}
}

The one-loop gluon self-energy is sum of the four
contributions shown in Fig.~(\ref{fig:gluonself}) plus an ultraviolet
subtraction,
\beqn
\mi\Pi^{\mu\nu\,R}_{ab}(p) &=&\int\frac{d^dl}{(2\pi)^d}
\left[\mi \widetilde{\Pi}^{\mu\nu}_{ab}(l,p)
-\mi \widetilde{\Pi}^{\mu\nu}_{ab\,UV}(l,p)\right]
\;\;.
\eeqn
Applying the Feynman rules, one finds that the integrand
$\widetilde{\Pi}^{\mu\nu}$ is
\beqn\nn\label{integrand:gluonself}
\widetilde{\Pi}^{\mu\nu}_{ab}(l,p) &=&  
-\mi\gs^2 \mu^{2\epsilon} \, C_A \delta_{ab}\, \frac{1}{2!}
\frac{V^{\mu\sigma\delta}(-p,-l_1,l_2)
  V^{\nu}_{\phantom{\mu}\delta\sigma}(p,-l_2,l_1) 
  - l_1^\mu l_2^\nu - l_2^\mu l_1^\nu}
{(l_1^2+\mi0)(l_2^2+\mi0)}\\
&&\;\;+\mi\,\gs^2 \mu^{2\epsilon} n_f \, T_R \delta_{ab}\,
\frac{\mathrm{Tr}\left[\gamma^\mu \s{l}_1\gamma^\nu \s{l}_2\right]}
{(l_1^2+\mi0)(l_2^2+\mi0)}\;\;,
\eeqn
where $a$, $b$ are the color indices,  $V^{\mu\sigma\delta}$ is the 
kinematical part of the three-gluon vertex, and the loop integral is
parametrized by
\beqn
l_1^\mu = l^\mu - \frac{1}{2} p^\mu\;\;, \qquad l_2^\mu = l^\mu 
+ \frac{1}{2} p^\mu\;\;.
\eeqn

The integrand of the gluon self-energy has quadratic and logarithmic
ultraviolet divergences. As pointed out in Sec.~\ref{sec:UV}, the
quadratic divergence requires a subtraction defined by a somewhat
different prescription than the simple prescription described in general
terms in Sec.~\ref{sec:UV} and used elsewhere in this appendix. We choose
the counterterm to be 
\beqn\nn
\widetilde{\Pi}^{\mu\nu}_{ab\, UV}(l,p) &=& 
-\mi\gs^2 \mu^{2\epsilon} \delta_{ab}
\frac{((d-2)C_A-4T_Rn_f)\left(2l^\mu l^\nu +
  \frac{1}{2d-2}(g^{\mu\nu}p^2-p^\mu p^\nu)\right)}
{\left(\left(l-\frac{1}{2}p\right)^2+\mi0\right)
\left(\left(l+\frac{1}{2}p\right)^2+\mi0\right)}\\\nn
&\phantom{=}&-\mi\gs^2 \mu^{2\epsilon} \delta_{ab}
\frac{(C_A+4T_Rn_f)\left(l^2 +\frac{1}{4} p^2\right)g^{\mu\nu}}
{\left(\left(l-\frac{1}{2}p\right)^2+\mi0\right)
\left(\left(l+\frac{1}{2}p\right)^2+\mi0\right)}\\
&\phantom{=}&-\mi\gs^2 \mu^{2\epsilon} \delta_{ab}
\left[\frac{3d-2}{2d-2}\frac{C_A(g^{\mu\nu}p^2-p^\mu p^\nu)}
    {(l^2 - \mu^2e^{1/15}+\mi0)^2}
 - \frac{2d-4}{d-1}\frac{T_R n_f(g^{\mu\nu}p^2-p^\mu p^\nu)}
    {(l^2 - \mu^2e^{-1/3}+\mi0)^2}\right] .\ \
\qquad\;
\eeqn
The first two terms eliminate all the quadratic divergences and the
integral of them is exactly zero in $d=4-2\epsilon$ dimensions. The last
term cancels the remaining logarithmic divergences. The integrand
$[\widetilde{\Pi}^{\mu\nu}_{ab}(l,p)
-\widetilde{\Pi}^{\mu\nu}_{ab\,UV}(l,p)]$ has the familiar tensor
structure $g^{\mu\nu}p^2 - p^\mu p^\nu$.

\subsection*{Quark-gluon vertex}

\FIGURE[ht]{
\hspace*{1.5cm}
\begin{picture}(120,120)(-60,-70)
  \Vertex(0,0){1.5}
  \Vertex(30,-30){1.5}
  \Vertex(-30,-30){1.5}
  \Gluon(0,0)(0,30){4}{4}
  \Gluon(30,-30)(-30,-30){3.5}{8}
  \ArrowLine(50,-50)(30,-30)
  \ArrowLine(30,-30)(0,0)
  \ArrowLine(0,0)(-30,-30)
  \ArrowLine(-30,-30)(-50,-50)
  \LongArrowArc(0,-17)(7,110,70)
  \Text(-53,-53)[t]{$p_2$, $\beta$}
  \Text(53,-53)[t]{$p_3$, $\alpha$}
  \Text(0,35)[]{$a$, $\mu$}
  \Text(15,15)[]{$p_1$}
  \Text(0,-40)[t]{$l_1$}
  \Text(20,-17.5)[bl]{$l_2$}
  \Text(-20,-17.5)[br]{$l_3$}
\end{picture}
\hfill
\begin{picture}(120,120)(-60,-70)
  \Vertex(0,0){1.5}
  \Vertex(30,-30){1.5}
  \Vertex(-30,-30){1.5}
  \Gluon(0,0)(0,30){4}{4}
  \ArrowLine(30,-30)(-30,-30)
  \ArrowLine(50,-50)(30,-30)
  \ArrowLine(-30,-30)(-50,-50)
  \Gluon(0,0)(30,-30){3.5}{5}
  \Gluon(-30,-30)(0,0){3.5}{5}
  \LongArrowArc(0,-19)(7,110,70)
  \Text(-53,-53)[t]{$p_2$, $\beta$}
  \Text(53,-53)[t]{$p_3$, $\alpha$}
  \Text(0,35)[]{$a$, $\mu$}
  \Text(15,15)[]{$p_1$}
  \Text(0,-40)[t]{$l_1$}
  \Text(23,-17.5)[bl]{$l_2$}
  \Text(-23,-17.5)[br]{$l_3$}
\end{picture}
\hspace*{1.5cm}
\caption{\label{vertex:qg}
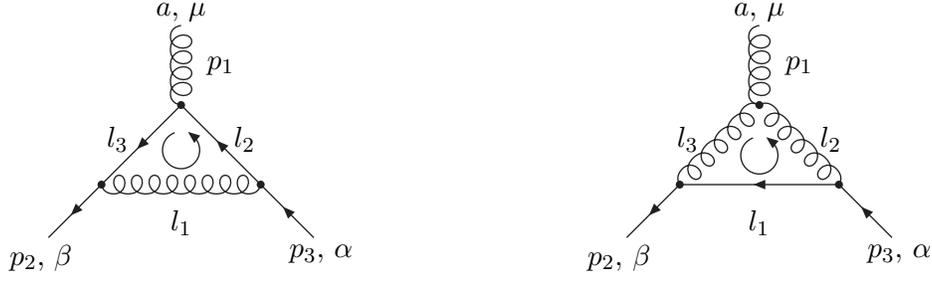
One-loop contributions to the quark-gluon vertex.}
}

At one-loop level there are two graphs that contribute to the quark-gluon
vertex, as shown in Fig.~\ref{vertex:qg}. These graphs have only
logarithmic ultraviolet divergences. The corresponding renormalized
quark-gluon vertex is   
\beqn\nn
\Gamma^{a\,R}_{\mu\beta\alpha}(p_1,p_2,p_3)&=& 
\mi\gs\mu^\epsilon t^a_{\beta\alpha} \gamma_\mu\\\nn
&+& \mi\gs\mu^\epsilon t^a_{\beta\alpha}\int \frac{d^dl}{(2\pi)^d}
\left[\widetilde{Q}_\mu(l, p_1,p_2,p_3) - \widetilde{Q}_\mu^{UV}(l)
\right]\\
&+& \mi\gs\mu^\epsilon t^a_{\beta\alpha}\int \frac{d^dl}{(2\pi)^d}
\left[\widetilde{G}_\mu(l, p_1,p_2,p_3)
  - \widetilde{G}_{\mu}^{UV}(l)
\right]\;\;.\qquad
\eeqn
The integrands $\widetilde{Q}_\mu$ and $\widetilde{G}_\mu$ are derived
directly from the Feynman rules,
\beqn
\widetilde{Q}_\mu(l, p_1,p_2,p_3) &=& 
-\mi\gs^2\mu^{2\epsilon}\left(C_F-\frac{C_A}2\right)
\frac{\gamma^\nu\s{l}_3\gamma_\mu\s{l}_2\gamma_\nu}
{(l_1^2+\mi0)(l_2^2+\mi0)(l_3^2+\mi0)}\;\;,\\
\widetilde{G}_\mu(l, p_1,p_2,p_3) &=& 
-\mi\gs^2\mu^{2\epsilon}\frac{C_A}2
\frac{\gamma^\nu\s{l}_1\gamma^\sigma V_{\mu\nu\sigma}(p_1,l_3,-l_2)}
{(l_1^2+\mi0)(l_2^2+\mi0)(l_3^2+\mi0)}\;\;,
\eeqn
where $V_{\mu\nu\sigma}$ is the kinematical part of the
triple gluon vertex and the loop integral was parametrized by
\beqn
l_1^\mu = l^\mu + \frac{p^\mu_3-p^\mu_2}3\;\;,\qquad 
l_2^\mu = l^\mu + \frac{p^\mu_1-p^\mu_3}3\;\;,\qquad 
l_3^\mu = l^\mu + \frac{p^\mu_2-p^\mu_1}3\;\;.
\eeqn
The corresponding ultraviolet counterterms are 
\beqn\label{vertex:qqg:UVcounter}
\widetilde{Q}^{UV}_{\mu}(l) &=& 
-\mi\gs^2\mu^{2\epsilon}(d-2) \left(C_F-\frac{C_A}2\right)
\frac{l^2\gamma_\mu - 2\s{l}l_\mu}
{(l^2-\mu^2e^{-2}+\mi0)^3}\;\;,\\
\widetilde{G}^{UV}_{\mu}(l) &=& 
-\mi\gs^2\mu^{2\epsilon}(d-2) \frac{C_A}2 
\frac{l^2\gamma_\mu(1+\epsilon)+2\s{l}l_\mu}
{(l^2-\mu^2e^{-2/3}+\mi0)^3}
\;\;.\qquad
\eeqn

The quark-photon vertex correction can be gotten from the quark-gluon
vertex with the substitutions $C_A \to 0$, $C_F \to 1$, $g \to -Q|e|$ and
$t^\alpha_{\beta\gamma} \to 1$, where $Q$ is the quark charge in units of
$|e|$. Thus the renormalized quark-photon  vertex is 
\beqn
\Gamma_\mu(p_1,p_2,p_3) = - \mi Q|e| \gamma_\mu
- \mi Q|e| \int \frac{d^dl}{(2\pi)^d}
\left[\widetilde{Q}_\mu(l, p_1,p_2,p_3)
  - \widetilde{Q}^{UV}_{\mu}(l)
\right]_{\substack{C_A=0 \\ C_F = 1}}
\;\;.
\eeqn

\subsection*{Three-gluon vertex}

The three-gluon vertex is sum of the five triangle graphs and three bubble
graphs shown in Fig.~\ref{vertex:ggg1}. 
\FIGURE[ht]{
\begin{picture}(120,120)(-60,-70)
  \Vertex(0,0){1.5}
  \Vertex(30,-30){1.5}
  \Vertex(-30,-30){1.5}
  \Gluon(0,0)(0,30){4}{4}
  \Gluon(-30,-30)(0,0){3.5}{6}
  \Gluon(0,0)(30,-30){3.5}{6}
  \Gluon(30,-30)(-30,-30){3.5}{9}
  \Gluon(-50,-50)(-30,-30){3.5}{4}
  \Gluon(30,-30)(50,-50){3.5}{4}
  \LongArrowArc(0,-17)(6,110,70)
  \Text(-48,-40)[br]{$p_2$}
  \Text(48,-40)[bl]{$p_3$}
  \Text(-53,-53)[t]{$a_2$, $\mu_2$}
  \Text(53,-53)[t]{$a_3$, $\mu_3$}
  \Text(0,35)[]{$a_1$, $\mu_1$}
  \Text(15,15)[]{$p_1$}
  \Text(0,-40)[t]{$l_1$}
  \Text(23,-17.5)[bl]{$l_2$}
  \Text(-23,-17.5)[br]{$l_3$}
\end{picture}
\hfill
\begin{picture}(120,120)(-60,-70)
  \Vertex(0,0){1.5}
  \Vertex(30,-30){1.5}
  \Vertex(-30,-30){1.5}
  \Gluon(0,0)(0,30){4}{4}
  \DashArrowLine(-30,-30)(0,0){1}
  \DashArrowLine(0,0)(30,-30){1}
  \DashArrowLine(30,-30)(-30,-30){1}
  \Gluon(-50,-50)(-30,-30){3.5}{4}
  \Gluon(30,-30)(50,-50){3.5}{4}
  \LongArrowArc(0,-17)(6,110,70)
  \Text(-48,-40)[br]{$p_2$}
  \Text(48,-40)[bl]{$p_3$}
  \Text(-53,-53)[t]{$a_2$, $\mu_2$}
  \Text(53,-53)[t]{$a_3$, $\mu_3$}
  \Text(0,35)[]{$a_1$, $\mu_1$}
  \Text(15,15)[]{$p_1$}
  \Text(0,-35)[t]{$l_1$}
  \Text(20,-17.5)[bl]{$l_2$}
  \Text(-20,-17.5)[br]{$l_3$}
\end{picture}
\hfill
\begin{picture}(120,120)(-60,-70)
  \Vertex(0,0){1.5}
  \Vertex(30,-30){1.5}
  \Vertex(-30,-30){1.5}
  \Gluon(0,0)(0,30){4}{4}
  \DashArrowLine(0,0)(-30,-30){1}
  \DashArrowLine(30,-30)(0,0){1}
  \DashArrowLine(-30,-30)(30,-30){1}
  \Gluon(-50,-50)(-30,-30){3.5}{4}
  \Gluon(30,-30)(50,-50){3.5}{4}
  \LongArrowArc(0,-17)(6,110,70)
  \Text(-48,-40)[br]{$p_2$}
  \Text(48,-40)[bl]{$p_3$}
  \Text(-53,-53)[t]{$a_2$, $\mu_2$}
  \Text(53,-53)[t]{$a_3$, $\mu_3$}
  \Text(0,35)[]{$a_1$, $\mu_1$}
  \Text(15,15)[]{$p_1$}
  \Text(0,-35)[t]{$l_1$}
  \Text(20,-17.5)[bl]{$l_2$}
  \Text(-20,-17.5)[br]{$l_3$}
\end{picture}
\\
$\begin{array}{c}
\begin{picture}(120,120)(-60,-70)
  \Vertex(0,0){1.5}
  \Vertex(30,-30){1.5}
  \Vertex(-30,-30){1.5}
  \Gluon(0,0)(0,30){4}{4}
  \ArrowLine(-30,-30)(0,0)
  \ArrowLine(0,0)(30,-30)
  \ArrowLine(30,-30)(-30,-30)
  \Gluon(-50,-50)(-30,-30){3.5}{4}
  \Gluon(30,-30)(50,-50){3.5}{4}
  \LongArrowArc(0,-17)(6,110,70)
  \Text(-48,-40)[br]{$p_2$}
  \Text(48,-40)[bl]{$p_3$}
  \Text(-53,-53)[t]{$a_2$, $\mu_2$}
  \Text(53,-53)[t]{$a_3$, $\mu_3$}
  \Text(0,35)[]{$a_1$, $\mu_1$}
  \Text(15,15)[]{$p_1$}
  \Text(0,-35)[t]{$l_1$}
  \Text(20,-17.5)[bl]{$l_2$}
  \Text(-20,-17.5)[br]{$l_3$}
\end{picture}
\end{array}$
\hfill
$\begin{array}{c}
\begin{picture}(120,120)(-60,-70)
  \Vertex(0,0){1.5}
  \Vertex(30,-30){1.5}
  \Vertex(-30,-30){1.5}
  \Gluon(0,0)(0,30){4}{4}
  \ArrowLine(0,0)(-30,-30)
  \ArrowLine(30,-30)(0,0)
  \ArrowLine(-30,-30)(30,-30)
  \Gluon(-50,-50)(-30,-30){3.5}{4}
  \Gluon(30,-30)(50,-50){3.5}{4}
  \LongArrowArc(0,-17)(6,110,70)
  \Text(-48,-40)[br]{$p_2$}
  \Text(48,-40)[bl]{$p_3$}
  \Text(-53,-53)[t]{$a_2$, $\mu_2$}
  \Text(53,-53)[t]{$a_3$, $\mu_3$}
  \Text(0,35)[]{$a_1$, $\mu_1$}
  \Text(15,15)[]{$p_1$}
  \Text(0,-35)[t]{$l_1$}
  \Text(20,-17.5)[bl]{$l_2$}
  \Text(-20,-17.5)[br]{$l_3$}
\end{picture}
\end{array}$
\hfill
$\dd \sum_{\{1,2,3\}'}\!\!\!\!\!\!\!\!
\begin{array}{c}
\begin{picture}(120,120)(-60,-70)
  \Vertex(0,3){1.5}
  \Vertex(0,-35){1.5}
  \Gluon(0,3)(0,30){4}{4}
  \GlueArc(0,-17.5)(17.5,-90,270){3}{16}
  \Gluon(0,-35)(-40,-50){3}{7}
  \Gluon(40,-50)(0,-35){3}{7}
  \LongArrowArc(0,-17)(9,110,70)
  \Text(-30,-40)[br]{$p_2$}
  \Text(30,-40)[bl]{$p_3$}
  \Text(-40,-53)[t]{$a_2$, $\mu_2$}
  \Text(40,-53)[t]{$a_3$, $\mu_3$}
  \Text(0,35)[]{$a_1$, $\mu_1$}
  \Text(15,15)[]{$p_1$}
  \Text(23,-17.5)[bl]{$l_1$}
  \Text(-23,-17.5)[br]{$l_2$}
\end{picture}
\end{array}$
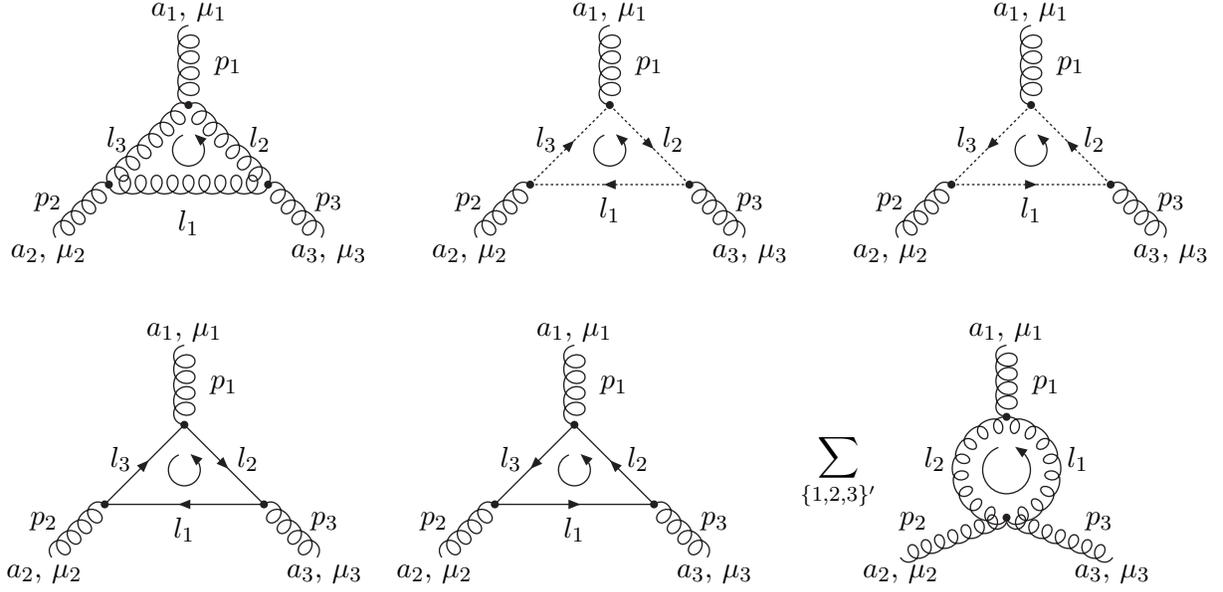
\caption{\label{vertex:ggg1}One-loop corrections to the three-gluon
  vertex. The $\{1,2,3\}'$ denotes the cyclic permutation of the indices
  $(1,2,3)$.}
}
It is useful to group the triangle graphs with a ghost loop together with
the triangle graphs with a gluon loop. The other graphs are treated
separately. Then the renormalized triple gluon vertex is
\beqn\nn
&&G^{\mu_1\mu_2\mu_3\,R}_{a_1a_2a_3}(p_1,p_2,p_3) = 
\mi\gs\mu^{2\epsilon} F^{a_1}_{a_2a_3}V^{\mu_1\mu_2\mu_3}(p_1,p_2,p_3)
\\\nn
&&\qquad +
\mi\gs\mu^{2\epsilon} F^{a_1}_{a_2a_3}\int \frac{d^dl}{(2\pi)^d}\left[ 
\widetilde{T}^{\mu_1\mu_2\mu_3}(l,p_1,p_2,p_3)
-\widetilde{T}_{UV}^{\mu_1\mu_2\mu_3}(l,p_1,p_2,p_3)
\right]\\
&&\qquad + 
\sum_{\{1,2,3\}'}\mi\gs\mu^{2\epsilon} F^{a_1}_{a_2a_3}
\int \frac{d^dl}{(2\pi)^d}\left[
\widetilde{B}^{\mu_1\mu_2\mu_3}(l,p_1)
-\widetilde{B}_{UV}^{\mu_1\mu_2\mu_3}(l,p_1)\right]
\;\;,\quad
\eeqn 
where $\widetilde{T}^{\mu_1\mu_2\mu_3}$ is the contribution of all
the triangle graphs, $\widetilde{B}^{\mu_1\mu_2\mu_3}$ represents
one of the bubble graphs, and $\{1,2,3\}'$ denotes the cyclic
permutations of the indices $(1,2,3)$.  The integrand of the triangle
graphs is given by
\beqn\nn
\widetilde{T}^{\mu_1\mu_2\mu_3}(l,p_1,p_2,p_3) &=& 
-\mi\gs^2\mu^{2\epsilon}\Bigg\{
\frac{C_A}2\,\frac{
{V^{\mu_1\delta} }_\lambda(p_1,l_3,-l_2)
{V^{\mu_2\gamma} }_\delta (p_2,l_1,-l_3)
{V^{\mu_3\lambda}}_\gamma (p_3,l_2,-l_1)
}{(l_1^2+\mi0)\,(l_2^2+\mi0)\,(l_3^2+\mi0)}\\\nn
&&\qquad\qquad\qquad
+\; \frac{C_A}2\,
\frac{l_3^{\mu_1} l_1^{\mu_2} l_2^{\mu_3}
  +l_2^{\mu_1} l_3^{\mu_2} l_1^{\mu_3}}
{(l_1^2+\mi0)\,(l_2^2+\mi0)\,(l_3^2+\mi0)}\\
&&\qquad\qquad\qquad
+\;T_R n_f\, \frac{\mathrm{Tr}\left[\gamma^{\mu_1} \s{l}_3\gamma^{\mu_2}
    \s{l}_1\gamma^{\mu_3}\s{l}_2\right]}
{(l_1^2+\mi0)\,(l_2^2+\mi0)\,(l_3^2+\mi0)}
\Bigg\}\;\;.
\eeqn
The loop integral here is parametrized as
\beqn
l_1^\mu = l^\mu + \frac{p^\mu_3-p^\mu_2}3\;\;,\qquad 
l_2^\mu = l^\mu + \frac{p^\mu_1-p^\mu_3}3\;\;,\qquad 
l_3^\mu = l^\mu + \frac{p^\mu_2-p^\mu_1}3\;\;.
\eeqn

With these choices, the ultraviolet counterterm
is given by
\beqn
\widetilde{T}_{UV}^{\mu_1\mu_2\mu_3}(l) &=& 
-\mi\gs^2\mu^{2\epsilon}\Bigg(\frac{C_A}2
\frac{N_g^{\mu_1\mu_2\mu_3}(l)}
{(l^2-\mu^2e^{-11/19}+\mi0)^3}
+T_R\,n_f\,\frac{N_q^{\mu_1\mu_2\mu_3}(l)}
{(l^2-\mu^2e^{-1}+\mi0)^3}\Bigg)\;\;,\qquad
\eeqn
where the numerator functions are 
\beqn\nn
N^{\mu_1\mu_2\mu_3}_g(l) &=& 
\sum_{\{1,2,3\}'}\left(
  -2 l^2 g^{\mu_1\mu_2}l^{\mu_3}
  - \frac83(d-2)l^{\mu_1} l^{\mu_2} l^{\mu_3}
  + \frac43(d-2) l^{\mu_1} l^{\mu_2} (p_2-p_1)^{\mu_3}
\right. \\
&&\qquad \qquad + 
\left.
  \frac{1}{3} g^{\mu_1\mu_2}
  \left(7\,l^2 (p_2-p_1)^{\mu_3}
    + 2\,l\!\cdot\!(p_2-p_1) l^{\mu_3} \right)
\right)
\eeqn
and 
\beqn\nn
N^{\mu_1\mu_2\mu_3}_q(l) &=& 
\sum_{\{1,2,3\}'}\left(
  -4 l^2 g^{\mu_1\mu_2}l^{\mu_3}
  +\frac{16}3 l^{\mu_1} l^{\mu_2} l^{\mu_3}
  - \frac83 l^{\mu_1} l^{\mu_2} (p_2-p_1)^{\mu_3}
\right. \\
&&\qquad\qquad + 
\left.
  \frac83 g^{\mu_1\mu_2}\left(l^2(p_2-p_1)^{\mu_3} -
    l\!\cdot\!(p_2-p_1)l^{\mu_3}\right)
\right)\;\;. 
\eeqn

The bubble integral is simpler. In this case the integrand has
only a logarithmic divergence. The integrand is given by
\beqn
\widetilde{B}^{\mu_1\mu_2\mu_3}(l_1,l_2, p_1) &=& 
\mi\gs^2\mu^{2\epsilon}\frac94C_A \frac{g^{\mu_1\mu_3}
  p_1^{\mu_2}-g^{\mu_1\mu_2} p_1^{\mu_3}}{(l_1^2+\mi0) (l_2^2+\mi0)}\;\;,
\eeqn
with the following parametrization for the loop integral,
\beqn
l_1^\mu = l^\mu + \frac{1}{2} p_1^\mu\;\;,\qquad
l_2^\mu = l^\mu - \frac{1}{2} p_1^\mu\;\;.
\eeqn
The corresponding counterterm for the bubble integrals is 
\beqn
\widetilde{B}_{UV}^{\mu_1\mu_2\mu_3}(l, p_1) &=& 
\mi\gs^2\mu^{2\epsilon}\frac94C_A \frac{g^{\mu_1\mu_3}
  p_1^{\mu_2}-g^{\mu_1\mu_2} p_1^{\mu_3}}{(l^2-\mu^2+\mi0)^2}\;\;.
\eeqn

\subsection*{Four-gluon vertex}

\FIGURE[ht]{
$\dd \sum_{P_1}\begin{array}{c}
\begin{picture}(120,120)(-60,-60)
  \Vertex(20,20){1.5}
  \Vertex(20,-20){1.5}
  \Vertex(-20,-20){1.5}
  \Vertex(-20,20){1.5}
  \Gluon(20,20)(20,-20){4}{5}
  \Gluon(20,-20)(-20,-20){4}{5}
  \Gluon(-20,-20)(-20,20){4}{5}
  \Gluon(-20,20)(20,20){4}{5}
  \Gluon(20,20)(40,40){4}{4}
  \Gluon(40,-40)(20,-20){4}{4}
  \Gluon(-20,-20)(-40,-40){4}{4}
  \Gluon(-40,40)(-20,20){4}{4}
  \LongArrowArc(0,0)(10,110,70)
  \Text(0,-30)[t]{$l_1$}
  \Text(0, 30)[b]{$l_3$}
  \Text(30, 0)[l]{$l_2$}
  \Text(-30, 0)[r]{$l_4$}
  \Text(-40,-45)[t]{$a_2$, $\mu_2$}
  \Text(40,-45)[t]{$a_3$, $\mu_3$}
  \Text(-40,45)[b]{$a_1$, $\mu_1$}
  \Text(40,45)[b]{$a_4$, $\mu_4$}
  \Text(-45,-40)[br]{$p_2$}
  \Text(45,-40)[bl]{$p_3$}
  \Text(-45,40)[tr]{$p_1$}
  \Text(45,40)[tl]{$p_4$}
\end{picture}
\end{array}
$
\hfill
$\dd \sum_{P_1}\begin{array}{c}
\begin{picture}(120,120)(-60,-60)
  \Vertex(20,20){1.5}
  \Vertex(20,-20){1.5}
  \Vertex(-20,-20){1.5}
  \Vertex(-20,20){1.5}
  \DashArrowLine( 20, 20)( 20,-20){1}
  \DashArrowLine( 20,-20)(-20,-20){1}
  \DashArrowLine(-20,-20)(-20, 20){1}
  \DashArrowLine(-20, 20)( 20, 20){1}
  \Gluon(20,20)(40,40){4}{4}
  \Gluon(40,-40)(20,-20){4}{4}
  \Gluon(-20,-20)(-40,-40){4}{4}
  \Gluon(-40,40)(-20,20){4}{4}
  \LongArrowArc(0,0)(10,110,70)
  \Text(0,-30)[t]{$l_1$}
  \Text(0, 30)[b]{$l_3$}
  \Text(30, 0)[l]{$l_2$}
  \Text(-30, 0)[r]{$l_4$}
  \Text(-40,-45)[t]{$a_2$, $\mu_2$}
  \Text(40,-45)[t]{$a_3$, $\mu_3$}
  \Text(-40,45)[b]{$a_1$, $\mu_1$}
  \Text(40,45)[b]{$a_4$, $\mu_4$}
  \Text(-45,-40)[br]{$p_2$}
  \Text(45,-40)[bl]{$p_3$}
  \Text(-45,40)[tr]{$p_1$}
  \Text(45,40)[tl]{$p_4$}
\end{picture}
\end{array}
$
\hfill
$\dd \sum_{P_1}\begin{array}{c}
\begin{picture}(120,120)(-60,-60)
  \Vertex(20,20){1.5}
  \Vertex(20,-20){1.5}
  \Vertex(-20,-20){1.5}
  \Vertex(-20,20){1.5}
  \DashArrowLine( 20,-20)( 20, 20){1}
  \DashArrowLine(-20,-20)( 20,-20){1}
  \DashArrowLine(-20, 20)(-20,-20){1}
  \DashArrowLine( 20, 20)(-20, 20){1}
  \Gluon(20,20)(40,40){4}{4}
  \Gluon(40,-40)(20,-20){4}{4}
  \Gluon(-20,-20)(-40,-40){4}{4}
  \Gluon(-40,40)(-20,20){4}{4}
  \LongArrowArc(0,0)(10,110,70)
  \Text(0,-30)[t]{$l_1$}
  \Text(0, 30)[b]{$l_3$}
  \Text(30, 0)[l]{$l_2$}
  \Text(-30, 0)[r]{$l_4$}
  \Text(-40,-45)[t]{$a_2$, $\mu_2$}
  \Text(40,-45)[t]{$a_3$, $\mu_3$}
  \Text(-40,45)[b]{$a_1$, $\mu_1$}
  \Text(40,45)[b]{$a_4$, $\mu_4$}
  \Text(-45,-40)[br]{$p_2$}
  \Text(45,-40)[bl]{$p_3$}
  \Text(-45,40)[tr]{$p_1$}
  \Text(45,40)[tl]{$p_4$}
\end{picture}
\end{array}
$
\\
$\dd \sum_{P_1}\!\!\begin{array}{c}
\begin{picture}(120,120)(-60,-60)
  \Vertex(20,20){1.5}
  \Vertex(20,-20){1.5}
  \Vertex(-20,-20){1.5}
  \Vertex(-20,20){1.5}
  \ArrowLine( 20, 20)( 20,-20)
  \ArrowLine( 20,-20)(-20,-20)
  \ArrowLine(-20,-20)(-20, 20)
  \ArrowLine(-20, 20)( 20, 20)
  \Gluon(20,20)(40,40){4}{4}
  \Gluon(40,-40)(20,-20){4}{4}
  \Gluon(-20,-20)(-40,-40){4}{4}
  \Gluon(-40,40)(-20,20){4}{4}
  \LongArrowArc(0,0)(10,110,70)
  \Text(0,-30)[t]{$l_1$}
  \Text(0, 30)[b]{$l_3$}
  \Text(30, 0)[l]{$l_2$}
  \Text(-30, 0)[r]{$l_4$}
  \Text(-40,-45)[t]{$a_2$, $\mu_2$}
  \Text(40,-45)[t]{$a_3$, $\mu_3$}
  \Text(-40,45)[b]{$a_1$, $\mu_1$}
  \Text(40,45)[b]{$a_4$, $\mu_4$}
  \Text(-45,-40)[br]{$p_2$}
  \Text(45,-40)[bl]{$p_3$}
  \Text(-45,40)[tr]{$p_1$}
  \Text(45,40)[tl]{$p_4$}
\end{picture}
\end{array}
$
\hfill
$\dd \sum_{P_1}\!\!\begin{array}{c}
\begin{picture}(120,120)(-60,-60)
  \Vertex(20,20){1.5}
  \Vertex(20,-20){1.5}
  \Vertex(-20,-20){1.5}
  \Vertex(-20,20){1.5}
  \ArrowLine( 20,-20)( 20, 20)
  \ArrowLine(-20,-20)( 20,-20)
  \ArrowLine(-20, 20)(-20,-20)
  \ArrowLine( 20, 20)(-20, 20)
  \Gluon(20,20)(40,40){4}{4}
  \Gluon(40,-40)(20,-20){4}{4}
  \Gluon(-20,-20)(-40,-40){4}{4}
  \Gluon(-40,40)(-20,20){4}{4}
  \LongArrowArc(0,0)(10,110,70)
  \Text(0,-30)[t]{$l_1$}
  \Text(0, 30)[b]{$l_3$}
  \Text(30, 0)[l]{$l_2$}
  \Text(-30, 0)[r]{$l_4$}
  \Text(-40,-45)[t]{$a_2$, $\mu_2$}
  \Text(40,-45)[t]{$a_3$, $\mu_3$}
  \Text(-40,45)[b]{$a_1$, $\mu_1$}
  \Text(40,45)[b]{$a_4$, $\mu_4$}
  \Text(-45,-40)[br]{$p_2$}
  \Text(45,-40)[bl]{$p_3$}
  \Text(-45,40)[tr]{$p_1$}
  \Text(45,40)[tl]{$p_4$}
\end{picture}
\end{array}
$
\hfill
$\dd \sum_{P_2}\begin{array}{c}
\begin{picture}(120,120)(-60,-60)
  \Vertex(20,20){1.5}
  \Vertex(20,-20){1.5}
  \Vertex(-20,0){1.5}
  \Gluon(20,20)(20,-20){4}{5}
  \Gluon(20,-20)(-20,0){4}{6}
  \Gluon(-20,0)(20,20){4}{6}

  \Gluon(20,20)(40,40){4}{4}
  \Gluon(40,-40)(20,-20){4}{4}
  \Gluon(-20,0)(-40,-40){4}{6}
  \Gluon(-40,40)(-20,0){4}{6}
  \LongArrowArc(6.2,0)(6.7,110,70)
  \Text(0,-20)[t]{$l_3$}
  \Text(0, 20)[b]{$l_2$}
  \Text(30, 0)[l]{$l_1$}
  \Text(-40,-45)[t]{$a_2$, $\mu_2$}
  \Text(40,-45)[t]{$a_3$, $\mu_3$}
  \Text(-40,45)[b]{$a_1$, $\mu_1$}
  \Text(40,45)[b]{$a_4$, $\mu_4$}
  \Text(-45,-40)[br]{$p_2$}
  \Text(45,-40)[bl]{$p_3$}
  \Text(-45,40)[tr]{$p_1$}
  \Text(45,40)[tl]{$p_4$}
\end{picture}
\end{array}
$
\\
$\dd \sum_{P_3}\begin{array}{c}
\begin{picture}(120,120)(-60,-60)
  \Vertex( 20,0){1.5}
  \Vertex(-20,0){1.5}
  \GlueArc(0,0)(20,0,180){4}{9}
  \GlueArc(0,0)(20,180,360){4}{9}
  \Gluon(40,40)(20,  0){4}{6}
  \Gluon(20, 0)(40,-40){4}{6}
  \Gluon(-40,-40)(-20, 0){4}{6}
  \Gluon(-20,  0)(-40,40){4}{6}
  \LongArrowArc(0,0)(10,110,70)
  \Text(0,-30)[t]{$l_1$}
  \Text(0, 30)[b]{$l_2$}
  \Text(-40,-45)[t]{$a_2$, $\mu_2$}
  \Text(40,-45)[t]{$a_3$, $\mu_3$}
  \Text(-40,45)[b]{$a_1$, $\mu_1$}
  \Text(40,45)[b]{$a_4$, $\mu_4$}
  \Text(-45,-40)[br]{$p_2$}
  \Text(45,-40)[bl]{$p_3$}
  \Text(-45,40)[tr]{$p_1$}
  \Text(45,40)[tl]{$p_4$}
\end{picture}
\end{array}
$
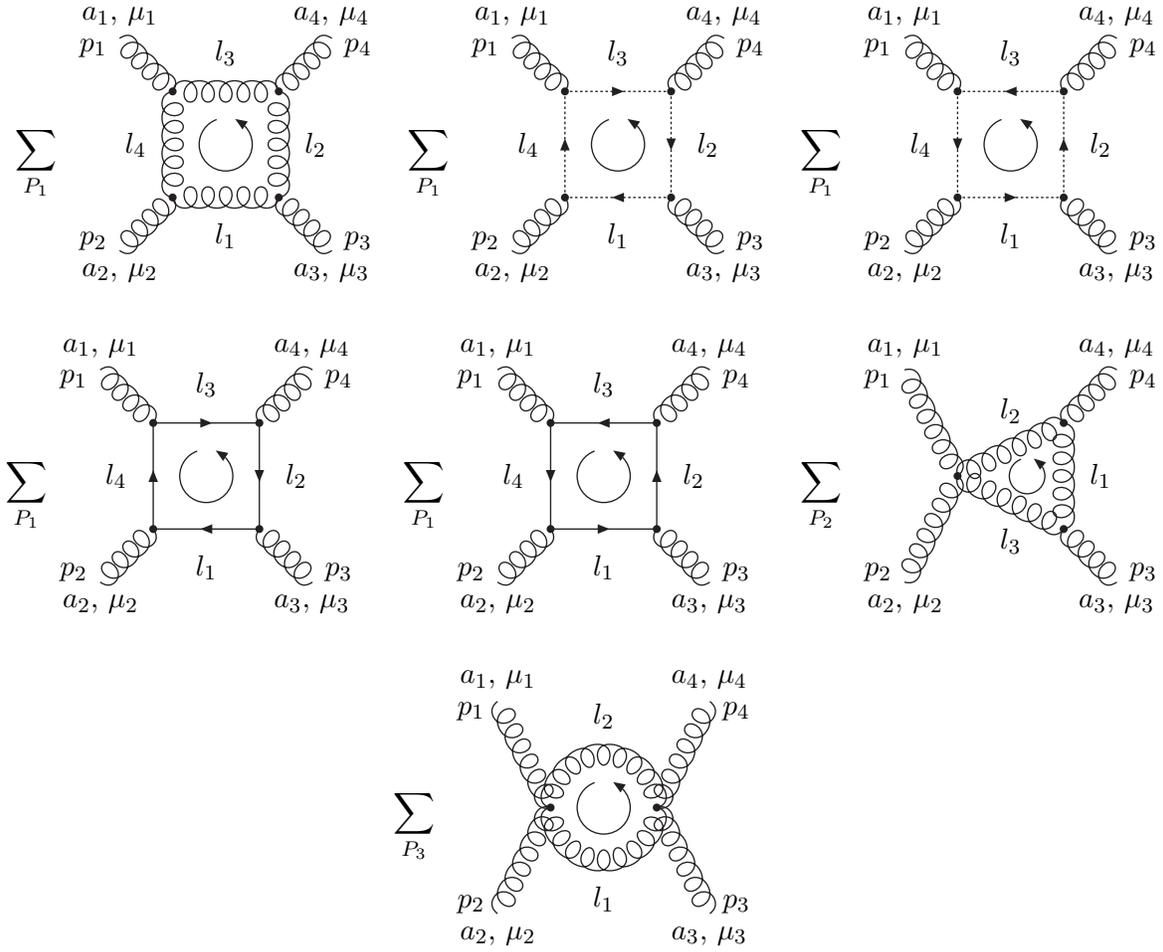
\caption{\label{fig:vertex4g}One-loop level correction to the
  four-point gluon vertex. The $P_1$, $P_2$, $P_3$ denote the
  permutations of the indices $(1,2,3,4)$ and they are given 
  in Eq.~(\ref{g4permut}).} 
}
The one-loop correction to the four-gluon vertex is sum of box,
triangle, and bubble graphs. In the case of the box graphs, it is
useful to group the graphs with a ghost loop together with the graphs
with a gluon loop. We write the renormalized vertex as     
\beqn\nn
&&Q^{\mu_1\mu_2\mu_3\mu_4}_{a_1a_2a_3a_4\,R}(p_1,p_2,p_3,p_4) = 
\mi \gs^2 \mu^{2\epsilon} W^{\mu_1\mu_2\mu_3\mu_4}_{a_1a_2a_3a_4}\\\nn
&&\qquad+ \mi \gs^2 \mu^{2\epsilon} \sum_{P_1}\int \frac{d^dl}{(2\pi)^d}
\left[\widetilde{B}^{\mu_1\mu_2\mu_3\mu_4}_{a_1a_2a_3a_4}
(l, p_1,p_2,p_3,p_4)
  - \widetilde{B}^{\mu_1\mu_2\mu_3\mu_4}_{UV\, a_1a_2a_3a_4}(l)
\right]\\\nn
&&\qquad+ \mi \gs^2 \mu^{2\epsilon} \sum_{P_2}\int \frac{d^dl}{(2\pi)^d}
\left[\widetilde{T}^{\mu_1\mu_2\mu_3\mu_4}_{a_1a_2a_3a_4}
(l, p_1,p_2,p_3,p_4)
  - \widetilde{T}^{\mu_1\mu_2\mu_3\mu_4}_{UV\, a_1a_2a_3a_4}(l)
\right]\\
&&\qquad+ \mi \gs^2 \mu^{2\epsilon} \sum_{P_3}
\int \frac{d^dl}{(2\pi)^d}
\left[\widetilde{F}^{\mu_1\mu_2\mu_3\mu_4}_{a_1a_2a_3a_4}
(l, p_1,p_2,p_3,p_4)
  - \widetilde{F}^{\mu_1\mu_2\mu_3\mu_4}_{UV\, a_1a_2a_3a_4}(l)
\right]\;\;,
\eeqn
where the $P_1$, $P_2$ and $P_3$ are certain sets of permutations of
the indices $(1,2,3,4)$, 
\beqn\nn\label{g4permut}
P_1 &=& \{(1,2,3,4),\; (1,2,4,3),\; (1,3,2,4)\}\;\;,\\\nn
P_2 &=& \{(1,2,3,4),\; (1,4,3,2),\; (1,3,2,4),\; (4,2,3,1),\;
(3,2,1,4),\; (4,3,2,1)\}\;\;,\\
P_3 &=& \{(1,2,3,4),\;(1,3,2,4),\;(1,4,3,2)\}\;\;.
\eeqn
The integrand of the box diagrams can be obtained by applying the
Feynman rules,
\beqn\nn
\widetilde{B}^{\mu_1\mu_2\mu_3\mu_4}_{a_1a_2a_3a_4}(l, p_1,p_2,p_3,p_4)
=&-&\mi\gs^2\mu^{2\epsilon}
\frac{\mathrm{Tr}\left(F^{a_1}F^{a_2}F^{a_3}F^{a_4}\right)
N_g^{\mu_1\mu_2\mu_3\mu_4}(l_1,l_2,l_3,l_4)}
{\left(l_1^2+\mi0\right)\left(l_2^2+\mi0\right)
  \left(l_3^2+\mi0\right)\left(l_4^2+\mi0\right)}\\
&+&\mi\gs^2\mu^{2\epsilon}
\frac{2n_f\,\mathrm{Tr}\left(t^{a_1}t^{a_2}t^{a_3}t^{a_4}\right)
  N_q^{\mu_1\mu_2\mu_3\mu_4}(l_1,l_2,l_3,l_4)}
{\left(l_1^2+\mi0\right)\left(l_2^2+\mi0\right)
  \left(l_3^2+\mi0\right)\left(l_4^2+\mi0\right)}\;\;,\ \ 
\eeqn
where the loop integral is parametrized by
\beqn
\begin{array}{c}\dd
l_1^\mu = l^\mu + \frac{2p^\mu_3+p^\mu_4-p^\mu_2}4\;\;,\qquad 
l_2^\mu = l^\mu - \frac{2p^\mu_3+p^\mu_2-p^\mu_4}4\;\;,\\ \dd
l_3^\mu = l^\mu + \frac{2p^\mu_1+p^\mu_2-p^\mu_4}4\;\;,\qquad 
l_4^\mu = l^\mu - \frac{2p^\mu_1+p^\mu_4-p^\mu_2}4\;\;,
\end{array}
\eeqn
and the numerator functions $N_g$ and $N_q$ are given by
\beqn\nn
N_g^{\mu_1\mu_2\mu_3\mu_4}(l_1,l_2,l_3,l_4) &=& 
V^{\mu_1\alpha}_{\phantom{\mu_1\alpha}\delta}(l_3-l_4,l_4,-l_3)
V^{\mu_2\beta }_{\phantom{\mu_2\beta }\alpha}(l_4-l_1,l_1,-l_4)\\\nn
&&\quad\times V^{\mu_3\gamma}_{\phantom{\mu_3\gamma}\beta }
(l_1-l_2,l_2,-l_1)
V^{\mu_4\delta}_{\phantom{\mu_4\delta}\gamma}(l_2-l_3,l_3,-l_2)\\
&-& l_4^{\mu_1}l_1^{\mu_2}l_2^{\mu_3}l_3^{\mu_4}
-l_3^{\mu_1}l_4^{\mu_2}l_1^{\mu_3}l_2^{\mu_4}\;\;,\\
N_q^{\mu_1\mu_2\mu_3\mu_4}(l_1,l_2,l_3,l_4) &=& 
\mathrm{Tr}\left[\gamma^{\mu_1}\s{l}_4\gamma^{\mu_2}\s{l}_1
  \gamma^{\mu_3}\s{l}_2\gamma^{\mu_4}\s{l}_3\right]\;\;.
\eeqn
The integral of the box graphs has only a logarithmic divergence.
The ultraviolet counterterm is 
\beqn\nn
\widetilde{B}^{\mu_1\mu_2\mu_3\mu_4}_{UV\,a_1a_2a_3a_4}(l)
=
&-&\mi\gs^2\mu^{2\epsilon}\frac{\mathrm{Tr}
\left(F^{a_1}F^{a_2}F^{a_3}F^{a_4}\right)
N_g^{\mu_1\mu_2\mu_3\mu_4}(l,l,l,l)}
{\left(l^2-\mu^2e^{-2/3}+\mi0\right)^4}\\\nn
&+&\mi\gs^2\mu^{2\epsilon}\frac{2n_f\,\mathrm{Tr}
\left(t^{a_1}t^{a_2}t^{a_3}t^{a_4}\right)
\left(N_q^{\mu_1\mu_2\mu_3\mu_4}(l,l,l,l)
  -32\,l^{\mu_1}l^{\mu_2}l^{\mu_3}l^{\mu_4}\right)}
{\left(l^2-\mu^2e^{-4/3}+\mi0\right)^4}\\
&+&\mi\gs^2\mu^{2\epsilon}\frac{64n_f\,\mathrm{Tr}
\left(t^{a_1}t^{a_2}t^{a_3}t^{a_4}\right)
l^{\mu_1}l^{\mu_2}l^{\mu_3}l^{\mu_4}}
{\left(l^2-\mu^2e^{-3/2}+\mi0\right)^4}\;\;.
\eeqn

The integrand of triangle graphs is
\beqn
\widetilde{T}^{\mu_1\mu_2\mu_3\mu_4}_{a_1a_2a_3a_4}(l,p_1,p_2,p_3,p_4)
= -\mi\gs^2\mu^{2\epsilon}
\frac{F^{a_3}_{ca}F^{a_4}_{bc} 
  W^{\mu_1\mu_2\alpha\beta}_{a_1a_2a\,b}
  V^{\mu_4}_{\phantom{\mu_4}\beta\gamma}(p_4,l_2,-l_1)
  V^{\mu_3\gamma}_{\phantom{\mu_3\gamma}\alpha}(p_3,l_1,-l_3)
}{\left(l_1^2+\mi0\right)\left(l_2^2+\mi0\right)\left(l_3^2+\mi0\right)}
,
\quad\quad
\eeqn
where the loop integral is parametrized by
\beqn
l_1^\mu = l^\mu + \frac{p^\mu_4-p^\mu_3}3\;\;,\qquad 
l_2^\mu = l^\mu + \frac{p^\mu_1+p^\mu_2-p^\mu_4}3\;\;,\qquad 
l_3^\mu = l^\mu + \frac{p^\mu_3-p^\mu_1-p^\mu_2}3\;\;.
\eeqn
With this parametrization the UV counterterm to the triangle graph is
\beqn
\widetilde{T}^{\mu_1\mu_2\mu_3\mu_4}_{UV\,a_1a_2a_3a_4}(l)
= -\mi\gs^2\mu^{2\epsilon}
\frac{F^{a_3}_{ca}F^{a_4}_{bc} 
  W^{\mu_1\mu_2\alpha\beta}_{a_1a_2a\,b}
  V^{\mu_4}_{\phantom{\mu_4}\beta\gamma}(0,l,-l)
  V^{\mu_3\gamma}_{\phantom{\mu_3\gamma}\alpha}(0,l,-l)
}{\left(l^2-\mu^2e^{-2/9}+\mi0\right)^3}\;\;.
\eeqn

Applying the Feynman rules, the integrand of the bubble graphs can be
found easily,
\beqn
\widetilde{F}^{\mu_1\mu_2\mu_3\mu_4}_{a_1a_2a_3a_4}(l,p_1,p_2,p_3,p_4)
= -\mi\gs^2\mu^{2\epsilon}\frac{1}{2!}
\frac{g_{\alpha\alpha'}
g_{\beta\beta'}W^{\mu_1\mu_2\alpha\beta}_{a_1a_2a\,b}
W^{\beta'\alpha'\mu_3\mu_4}_{b\,a\,a_3a_4}
}{\left(l_1^2+\mi0\right)\left(l_2^2+\mi0\right)}\;\;,
\eeqn
where 
\beqn
l_1^\mu = l^\mu-\frac{p^\mu_1+p^\mu_2}2\;\;,\qquad
l_2^\mu = l^\mu+\frac{p^\mu_1+p^\mu_2}2\;\;.
\eeqn
With this, the UV counterterm to a bubble graph is
\beqn
\widetilde{F}^{\mu_1\mu_2\mu_3\mu_4}_{UV\,a_1a_2a_3a_4}(l)
= -\mi\gs^2\mu^{2\epsilon}\frac{1}{2!}
\frac{g_{\alpha\alpha'}g_{\beta\beta'}
W^{\mu_1\mu_2\alpha\beta}_{a_1a_2a\,b}
W^{\beta'\alpha'\mu_3\mu_4}_{b\,a\,a_3a_4}
}{\left(l^2-\mu^2e^{1/3}+\mi0\right)^2}\;\;.
\eeqn

\bibliography{onefile}

\providecommand{\href}[2]{#2}\begingroup\raggedright\begin{thebibliography}{10}

\bibitem{Ellis:1981wv}
R.~K. Ellis, D.~A. Ross, and A.~E. Terrano, {\it The perturbative calculation
  of jet structure in $e^+e^-$ annihilation},  {\em Nucl. Phys.} {\bf B178}
  (1981) 421.

\bibitem{Ellis:1986er}
R.~K. Ellis and J.~C. Sexton, {\it {QCD} radiative corrections to parton parton
  scattering},  {\em Nucl. Phys.} {\bf B269} (1986) 445.

\bibitem{Bern:1993mq}
Z.~Bern, L.~J. Dixon, and D.~A. Kosower, {\it One loop corrections to five
  gluon amplitudes},  {\em Phys. Rev. Lett.} {\bf 70} (1993) 2677--2680,
  [\href{http://xxx.lanl.gov/abs/hep-ph/9302280}{{\tt hep-ph/9302280}}].

\bibitem{Bern:1995fz}
Z.~Bern, L.~J. Dixon, and D.~A. Kosower, {\it One loop corrections to two quark
  three gluon amplitudes},  {\em Nucl. Phys.} {\bf B437} (1995) 259--304,
  [\href{http://xxx.lanl.gov/abs/hep-ph/9409393}{{\tt hep-ph/9409393}}].

\bibitem{Bern:1997ka}
Z.~Bern, L.~J. Dixon, D.~A. Kosower, and S.~Weinzierl, {\it One-loop amplitudes
  for $e^+e^- \to \bar{q}q\bar{Q}q$},  {\em Nucl. Phys.} {\bf B489} (1997)
  3--23, [\href{http://xxx.lanl.gov/abs/hep-ph/9610370}{{\tt hep-ph/9610370}}].

\bibitem{Bern:1998sc}
Z.~Bern, L.~J. Dixon, and D.~A. Kosower, {\it One-loop amplitudes for $e^+e^-$
  to four partons},  {\em Nucl. Phys.} {\bf B513} (1998) 3--86,
  [\href{http://xxx.lanl.gov/abs/hep-ph/9708239}{{\tt hep-ph/9708239}}].

\bibitem{Kunszt:1994tq}
Z.~Kunszt, A.~Signer, and Z.~Tr\'ocs\'anyi, {\it One loop radiative corrections
  to the helicity amplitudes of {QCD} processes involving four quarks and one
  gluon},  {\em Phys. Lett.} {\bf B336} (1994) 529--536,
  [\href{http://xxx.lanl.gov/abs/hep-ph/9405386}{{\tt hep-ph/9405386}}].

\bibitem{Soper:1998ye}
D.~E. Soper, {\it {QCD} calculations by numerical integration},  {\em Phys.
  Rev. Lett.} {\bf 81} (1998) 2638--2641,
  [\href{http://xxx.lanl.gov/abs/hep-ph/9804454}{{\tt hep-ph/9804454}}].

\bibitem{Soper:1999xk}
D.~E. Soper, {\it Techniques for {QCD} calculations by numerical integration},
  {\em Phys. Rev.} {\bf D62} (2000) 014009,
  [\href{http://xxx.lanl.gov/abs/hep-ph/9910292}{{\tt hep-ph/9910292}}].

\bibitem{Soper:2001hu}
D.~E. Soper, {\it Choosing integration points for {QCD} calculations by
  numerical integration},  {\em Phys. Rev.} {\bf D64} (2001) 034018,
  [\href{http://xxx.lanl.gov/abs/hep-ph/0103262}{{\tt hep-ph/0103262}}].

\bibitem{Kramer:2002cd}
M.~Kramer and D.~E. Soper, {\it Next-to-leading order numerical calculations in
  {Coulomb} gauge},  {\em Phys. Rev.} {\bf D66} (2002) 054017,
  [\href{http://xxx.lanl.gov/abs/hep-ph/0204113}{{\tt hep-ph/0204113}}].

\bibitem{Binoth:2000ps}
T.~Binoth and G.~Heinrich, {\it An automatized algorithm to compute infrared
  divergent multi-loop integrals},  {\em Nucl. Phys.} {\bf B585} (2000)
  741--759, [\href{http://xxx.lanl.gov/abs/hep-ph/0004013}{{\tt
  hep-ph/0004013}}].

\bibitem{Binoth:2002xh}
T.~Binoth, G.~Heinrich, and N.~Kauer, {\it A numerical evaluation of the scalar
  hexagon integral in the physical region},  {\em Nucl. Phys.} {\bf B654}
  (2003) 277--300, [\href{http://xxx.lanl.gov/abs/hep-ph/0210023}{{\tt
  hep-ph/0210023}}].

\bibitem{Binoth:2003ak}
T.~Binoth and G.~Heinrich, {\it Numerical evaluation of multi-loop integrals by
  sector decomposition},  \href{http://xxx.lanl.gov/abs/hep-ph/0305234}{{\tt
  hep-ph/0305234}}.

\bibitem{Catani:1997vz}
S.~Catani and M.~H. Seymour, {\it A general algorithm for calculating jet cross
  sections in {NLO} {QCD}},  {\em Nucl. Phys.} {\bf B485} (1997) 291--419,
  [\href{http://xxx.lanl.gov/abs/hep-ph/9605323}{{\tt hep-ph/9605323}}].

\bibitem{Gunion:1985vc}
J.~F. Gunion and Z.~Kunszt, {\it Improved analytic techniques for tree graph
  calculations and the $g g q \bar{q}$ lepton anti-lepton subprocess},  {\em
  Phys. Lett.} {\bf B161} (1985) 333.

\bibitem{Mangano:1991by}
M.~L. Mangano and S.~J. Parke, {\it Multiparton amplitudes in gauge theories},
  {\em Phys. Rept.} {\bf 200} (1991) 301--367.

\bibitem{Grammer:1973db}
G.~Grammer~Jr. and D.~R. Yennie, {\it Improved treatment for the infrared
  divergence problem in quantum electrodynamics},  {\em Phys. Rev.} {\bf D8}
  (1973) 4332--4344.

\bibitem{Giele:1992vf}
W.~T. Giele and E.~W.~N. Glover, {\it Higher order corrections to jet
  cross-sections in e+ e- annihilation},  {\em Phys. Rev.} {\bf D46} (1992)
  1980--2010.

\bibitem{Collins:1985ue}
J.~C. Collins, D.~E. Soper, and G.~Sterman, {\it Factorization for short
  distance hadron - hadron scattering},  {\em Nucl. Phys.} {\bf B261} (1985)
  104.

\end{thebibliography}\endgroup
\end{document}